\documentclass[10pt,preprint,floatfix,superscriptaddress,preprintnumbers,amsmath,amssymb,prb,aps,longbibliography]{revtex4-2}
\usepackage{graphicx,bm}
\usepackage{amssymb,amsmath}
\usepackage{float}
\usepackage{xcolor}
\usepackage{hyperref}
\hypersetup{colorlinks=true,linkcolor=red,citecolor=blue}
\DeclareUnicodeCharacter{FB01}{fi}
\usepackage[normalem]{ulem}

\begin{document}

\title{Giant topological and planar Hall effect in Cr$_{1/3}$NbS$_{2}$}

\author{D. A. Mayoh}
\email[]{d.mayoh.1@warwick.ac.uk}
\author{J. Bouaziz}
\author{A. E. Hall}
\author{J. B. Staunton}
\author{M. R. Lees}
\author{G. Balakrishnan}
\email[]{g.balakrishnan@warwick.ac.uk}
\affiliation{Physics Department, University of Warwick, Coventry, CV4 7AL, United Kingdom}

\begin{abstract}

Cr$_{1/3}$NbS$_{2}$ is a transition metal dichalcogenide that has been of significant interest due to its ability to host a magnetic chiral soliton lattice. Conventional and planar Hall measurements provide valuable insight into the detection of exotic spin structures in chiral magnets. We show that the presence of a giant planar Hall effect can be attributed to a tilted soliton lattice in Cr$_{1/3}$NbS$_{2}$. Our detailed angular dependent study shows the planar Hall effect and anisotropic magnetoresistance are intrinsically linked in complex non-coplanar magnets. From the conventional Hall signal we show the presence of a giant unconventional, likely topological Hall component, that is the fingerprint of non-coplanar spin textures.
\end{abstract}

\maketitle

\section{Introduction}

The fundamental characteristics of two-dimensional (2D) magnetic materials can be scrutinized by examining the behavior of their electrons in magnetic and electric fields. 2D magnetic materials have been found to be highly tunable in terms of their physical characteristics making them excellent candidates for the study of exotic quantum phases and for spintronic devices~\cite{Ningrum:2020,Mak:2019}. Chiral helimagnetism has provided an exciting twist in the story of 2D and bulk magnetic materials with various examples of self-localized spin textures having been reported experimentally including skyrmions~\cite{Muhlbauer:2009, Karube:2016, Kezsmarki:2015}, anti-skyrmions~\cite{Saha:2019}, bubbles~\cite{Liu:2016} and vortices~\cite{Pribaig:2007}. These magnetic states often emerge due to the Dzyaloshinskii-Moriya interaction (DMI)~\cite{Dzyaloshinky:1958, Moriya:1960} but other stabilization mechanisms can also be at play, such as multi-spin interactions~\cite{Heinze:2011spontaneous,Ozawa:2017,grytsiuk2020topological} or geometric frustration~\cite{Okubo:2012,Leonov:2015multiply}.

Hall measurements have proved to be an excellent tool for probing spin textures~\cite{Kanazawa:2011, Nakazawa:2018, Taguchi:2001, Spencer:2018}. Indeed, the hallmark of skyrmion-like materials is the so called topological Hall effect (THE)~\cite{Neubauer:2009, Porter:2014, Kurumaji:2019}. However, other forms of non-coplanar spin textures may also result in a finite THE. In particular, non-coplanar phases occurring in metallic compounds can cause variations in the magnetization which may lead to an emergent magnetic field that can deflect the electrons in the transverse direction~\cite{Schulz:2012}. A giant topological Hall coefficient is normally defined as $\rho^{\mathrm{THE}}_{\mathrm{xy}} \geq 1~\mu\Omega$~cm and has been reported in several compounds including Gd$_{2}$PdSi$_{3}$~\cite{Kurumaji:2019},  NdMn$_{2}$Ge$_{2}$~\cite{Wang:2020} and thin film oxides~\cite{Vistoli:2019giant}.

Cr-intercalated NbS$_{2}$ (Cr$_{1/3}$NbS$_{2}$) has attracted considerable attention due to the presence of a chiral helimagnetic state and, in an applied field, the observation of a chiral soliton lattice (CSL) by Lorentz transmission electron microscopy (LTEM)~\cite{Togawa:2012}. Cr$_{1/3}$NbS$_{2}$ is a member of the intercalated transition metal dichalcogenide family and has a chiral noncentrosymmetric hexagonal structure (space group: $P6_{3}22$)~\cite{Parkin:1980a, Parkin:1980b}. Cr$_{1/3}$NbS$_{2}$ consists of layers of Nb and S with Cr intercalated between the layers as shown in Fig.~\ref{FIG: Crystal structure}(a). It is known to have a complex magnetic phase diagram and five distinct magnetic phases have been detected through a variety of experimental techniques including LTEM, muon spin rotation/relaxation, magnetization, magnetocaloric effects, and magneto-transport~\cite{Cao:2020}. The magnetic ordering observed in Cr$_{1/3}$NbS$_{2}$, is highly dependent on the experimental conditions. In zero-applied field, a twisting array of magnetic moments in the $ab$-plane of Cr$_{1/3}$NbS$_{2}$ results in chiral helimagnetic (CHM) ordering~\cite{Togawa:2012,Togawa:2015}. Applying a magnetic field along the screw axis (the $c$-axis) results in the magnet moments being tilted towards the direction of the field giving a chiral conical (CC) phase~\cite{Yonemure:2017}, while increasing the magnetic field still further results in a forced ferromagnetism (FFM). If the magnetic field is applied perpendicular to the screw axis, a CSL is established. This lattice is formed of areas of forced ferromagnetism broken by a twist in the magnetic field~\cite{Togawa:2012}. Recently, torque magnetometry has revealed a tilted CSL phase in Cr$_{1/3}$NbS$_{2}$ when the magnetic field is applied at an angle greater than 1.5~degrees from the $c$-axis~\cite{Yonemure:2017}.

The key finding of this work is the presence of a giant THE in Cr$_{1/3}$NbS$_{2}$ that is the finger print of non-coplanar spin textures in this material. The emergence of a finite PHE is attributed to the presence of the tilted CSL. The temperature and field dependence of the PHE has been thoroughly investigated in parallel with its longitudinal counterpart, the anisotropic magnetoresistance (AMR). Both are found to be intimately linked, highlighting that the connection between the PHE and AMR is still valid for complex non-coplanar magnets. Furthermore, by disentangling the ordinary and anomalous Hall components from the conventional Hall data, we reveal the presence of a giant topological Hall effect. We attribute the origin of the topological Hall effect in Cr$_{1/3}$NbS$_{2}$ to the complex magnetic textures present in this material. \\



\section{Experimental Details}

Single crystals of Cr$_{1/3}$NbS$_{2}$ were produced by the chemical vapor transport technique using iodine as the transport agent. Polycrystalline powders of Cr$_{1/3}$NbS$_{2}$ along with 5~mg/cc of the transport agent were placed in an evacuated and sealed quartz tube. The tube was then heated with one end held at $950~^{\circ}$C and the other at $800~^{\circ}$C for three weeks, before cooling to room temperature. Several single crystals in the form of hexagonal platelets with sides $\sim 2$~mm in length were obtained in each growth. DC magnetization, $M$, measurements were performed as a function of temperature (1.8 to 300~K) and field (0 to 7~T) in a Quantum Design Magnetic Property Measurement System. Magnetic hysteresis loops were acquired using an Oxford Instruments vibrating sample magnetometer in magnetic fields up to 6~T. Resistivity, conventional Hall and planar Hall effect measurements were performed as a function of temperature (1.8 to 300~K) and field (0 to 9~T) using the ac transport option in a Quantum Design Physical Property Measurement System. Currents of between 1 and 10~mA at a frequency of 113~Hz were used for all the measurements. Further details on the fitting procedure used for the conventional Hall effect data can be found in Appendix~\ref{APDX: Hall effect}



\section{Magnetization and Longitudinal Resistivity.}

A single crystal of Cr$_{1/3}$NbS$_{2}$ with a $T_{\mathrm{c}}$ of 111~K was chosen for these experiments as shown in Fig.~\ref{FIG: Crystal structure}(b). A Laue diffraction pattern of the crystal is shown in the Supplemental Material~\cite{SMnote}. DC magnetization, $M$, versus temperature curves for the single crystal of Cr$_{1/3}$NbS$_{2}$ in various fields up to 7~T applied perpendicular to the platelet are shown in Fig.~\ref{FIG: Mag and Res}(c). In low fields the system exhibits a sharp transition to a magnetically ordered state at $T_{\mathrm{c}}=111$~K. As the field is increased the magnetic transition broadens. In 7~T we observe a significant magnetization up to 250~K. Magnetization as a function of applied field, $H$, where the field was applied nominally along the $c$~axis is shown in Fig.~\ref{FIG: Mag and Res}(d). At low-temperature the magnetization increases with increasing applied field until saturation at $M = 2.9(1)~\mu_{\mathrm{B}}/\mathrm{Cr~atom}$. The saturation magnetization of Cr$_{1/3}$NbS$_{2}$ is expected to be approximately $3~\mu_{\mathrm{B}}/\mathrm{Cr~atom}$ for chromium in the trivalent $(\mathrm{S}=3/2)$ state. The magnetization curves can be explained by the presence of the CHM slowly being polarized along the direction of the applied field. No significant hysteresis is visible in the $M(H)$ curves at any temperature. 

The different magnetic regimes in Cr$_{1/3}$NbS$_{2}$ are observed to have distinct electronic transport properties. The temperature dependence of the zero-field longitudinal resistivity $\rho_{\mathrm{xx}}$ is shown in Fig.~\ref{FIG: Mag and Res}(e). Also shown in Fig.~\ref{FIG: Mag and Res}(f--i) is $\rho_{\mathrm{xx}}$ vs temperature for assorted magnetic fields applied parallel to the $c$~axis. The temperature dependence of the longitudinal resistivity differs from a conventional metallic ferromagnet~\cite{Colvin:1960}. $\rho_{\mathrm{xx}}$ initially increases with decreasing temperature until $\sim 150$~K. A similar increase in $\rho_{\mathrm{xx}}$ is observed in UPd$_2$Si$_2$ and was attributed to the Kondo effect~\cite{Barati:1993}. Below 150~K the resistivity of Cr$_{1/3}$NbS$_{2}$ then falls over a wide temperature interval, $\Delta T = 100$~K, centered around $T_{\mathrm{c}}$. The drop in the resistivity at temperatures below the magnetic transition is due to the magnetic ordering which diminishes the spin-disorder scattering~\cite{Haas:1968}. This reduction is not monotonous but instead tends to saturate as observed in previous work~\cite{Ghimire:2012}. A possible origin of this behavior is a disorder of the Cr atoms~\cite{Dyadkin:2015}. However, given the presence of a helical phase/CSL in this temperature regime, there is no one to one correspondence between the behavior seen here and that observed in a simple ferromagnet. The residual resistivity at 2~K is $\rho_{\mathrm{xx},0} = 0.339(5)$~m$\Omega$~cm. 

The field dependence of $\rho_{\mathrm{xx}}$ in Cr$_{1/3}$NbS$_{2}$ for a selection of temperatures is shown in Figs.~\ref{FIG: Mag and Res}(e-h). Different behaviors are observed in the various temperature regimes dependent on the magnetic ordering. Between 2 and 20~K $\rho_{\mathrm{xx}}$ is seen to saturate in fields above 2~T. The rapid decrease in the resistivity from 0 to 2~T can be attributed to the presence of the CSL, and as the field is increased the moments are canted towards the direction of field which in turn diminishes the spin-disorder scattering. $\rho_{\mathrm{xx}}$ saturates when the spins are fully polarized along $H$. This field dependence of the magnetoresistance is typical of helimagnetic systems~\cite{Lee:2007,Chapman:2013}. A plateau in the magnetoresistance has also been attributed to a skyrmion lattice phase in Gd$_{2}$PdSi$_{3}$~\cite{Zhang:2020}. Between 20 and 110~K, the distinct change in slope of $\rho_{\mathrm{xx}}\left(H\right)$ around 2~T becomes more diffuse as the mean free path of the electrons decreases, therefore, they experience less of the CSL phase. At temperatures above 120~K, which is above $T_{\mathrm{c}}$, $\rho_{\mathrm{xx}}$ is seen to have parabolic dependence, typical of ferromagnetic materials, before becoming almost linear with $H$ at $300$~K. \\


\section{Hall effect in Cr$_{1/3}$NbS$_{2}$}.

\begin{figure}[tb!]
\centering
\includegraphics[width=0.5\columnwidth]{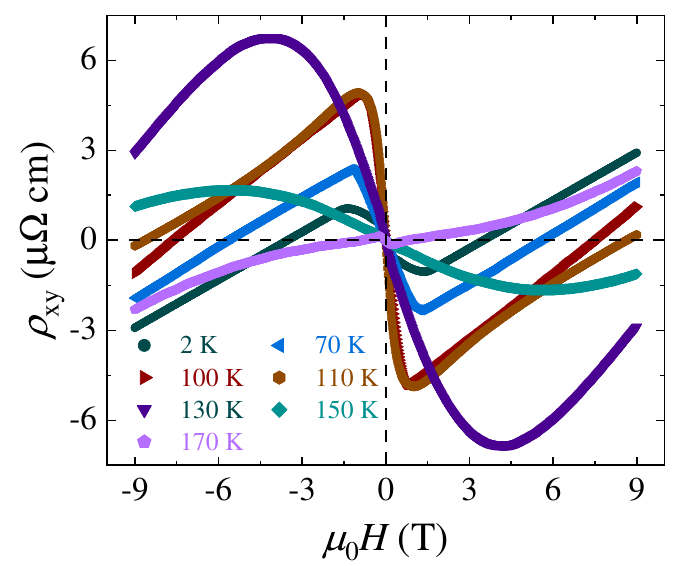}
\caption{(Color online) Field-dependent Hall resistivity $\rho_{\mathrm{xy}}$ at  10, 70, 110, 130, 190 and 230~K. There is a significant change in the behavior of the Hall effect as the temperature is lowered through $T_{\mathrm{c}}$. It is observed that the shape of the field dependence in Hall effect changes from a smooth almost parabolic shape to a sharp triangular shape.}
\label{FIG: Hall effect}
\end{figure}

Hall effect measurements on Cr$_{1/3}$NbS$_{2}$ were performed at a variety of temperatures, where the field was applied perpendicular to the largest face of the crystal platelet (nominally along the $c$~axis) and the current applied within the plane of the plate (nominally the $ab$~plane), and are shown in Fig.~\ref{FIG: Hall effect}. To understand the underlying physics in Cr$_{1/3}$NbS$_{2}$ we investigate the Hall response and disentangle its different components. The Hall signal is usually made up of several contributions:
\begin{equation}
\rho_{xy} = \rho^\text{OHE}_{xy} + \rho^\text{AHE}_{xy} + \rho^\text{THE}_{xy}. 
\label{EQ: Hall effect}
\end{equation}
\noindent The first term $\rho_{\mathrm{xy}}^{\mathrm{OHE}}$ is the ordinary Hall effect (OHE) which scales linearly with the magnetic field $B$, $\rho_{\mathrm{xy}}^{\mathrm{OHE}} = R_{\mathrm{H}}B$, where $R_{\mathrm{H}}$ is the Hall coefficient. The second term $\rho_{\mathrm{xy}}^{\mathrm{AHE}}$ is the anomalous Hall effect (AHE), which emerges from two mechanisms, one intrinsic and the other extrinsic. The intrinsic contribution is related to the presence of a non-vanishing Berry curvature leading to an internal magnetic field which deflects the electrons in a direction perpendicular to the applied electric field~\cite{Chang:1996,Nagaosa:2010review}. The extrinsic contribution can be related to side jump and skew scattering~\cite{Nagaosa:2010review, Smit:1955, Berger:1970}. Usually, the skew scattering contribution is the weakest component, however, it has been shown recently that it can have a measurable impact on the Hall response in some materials~\cite{Ernst:2019}. The third term, $\rho_{\mathrm{xy}}^{\mathrm{THE}}$, is the topological Hall effect~\cite{Taguchi:2001,Bruno:2004,Neubauer:2009} whose dependence on the magnetization is non-trivial, $\rho_{\mathrm{xy}}^{\mathrm{THE}}$ is proportional to the emergent field $\vec{B}_\text{e}$ created by a non-coplanar magnetic texture~\cite{Schulz:2012,Bouaziz:2021}, where $\vec{B}_{\mathrm{e}} \propto \vec{m} \cdot \left(\frac{\partial \vec{m}}{\partial x} \times \frac{\partial \vec{m}}{\partial y} \right)$. $\rho_{\mathrm{xy}}^{\mathrm{THE}}$ has a non-linear evolution with the magnetic field, and is only seen in the presence of non-coplanar phases~\cite{Schulz:2012}. This component of the Hall signal is anti-symmetric when the direction of the applied magnetic field is reversed~\cite{Vistoli:2019giant}.  At this point it worth considering whether any of the magnetic phases in Cr$_{1/3}$NbS$_{2}$ can generate a topological Hall contribution. Neither the helimagnetic phase nor the chiral soliton lattice will display a topological Hall effect since the scalar chirality must vanish, i.e. $\mathcal{C}_{ijk}= \vec{m}_{i}\cdot(\vec{m}_{j}\times\vec{m}_{k})$. However, for the case of a tilted-CSL~\cite{Yonemure:2017}, a finite topological Hall effect contribution is possible (see Appendix.~\ref{APDX: THE}).

The presence of tilted chiral soliton lattice may also lead to the appearance of a planar Hall effect in Cr$_{1/3}$NbS$_{2}$. The planar Hall effect emerges solely from the anisotropic magnetoresistance~\cite{Jan:1957,Seemann:2011,Juba:2016}, and occurs when the in-plane component of the magnetization is misaligned with respect to the applied current. For a current along the $x$~direction, $\rho^\text{PHE}_{xy}$ is ~\cite{Juba:2016,Seemann:2011,li:2020}: 
\begin{equation}
\rho^\text{PHE}_{xy} = (\rho_{\parallel}-\rho_{\perp})\,m_x\,m_y,
\label{PHE_func}
\end{equation}
where $m_x$ and $m_y$ are the in-plane components of the magnetization and $\rho_{\parallel}$ $(\rho_{\perp})$ are the longitudinal resistivity when $\vec{m}$ is parallel (perpendicular) to the current direction. \\


\section{Planar Hall effect and anisotropic magnetoresistance} 

\begin{figure*}[tb!]
\centering
\includegraphics[width=0.9\textwidth]{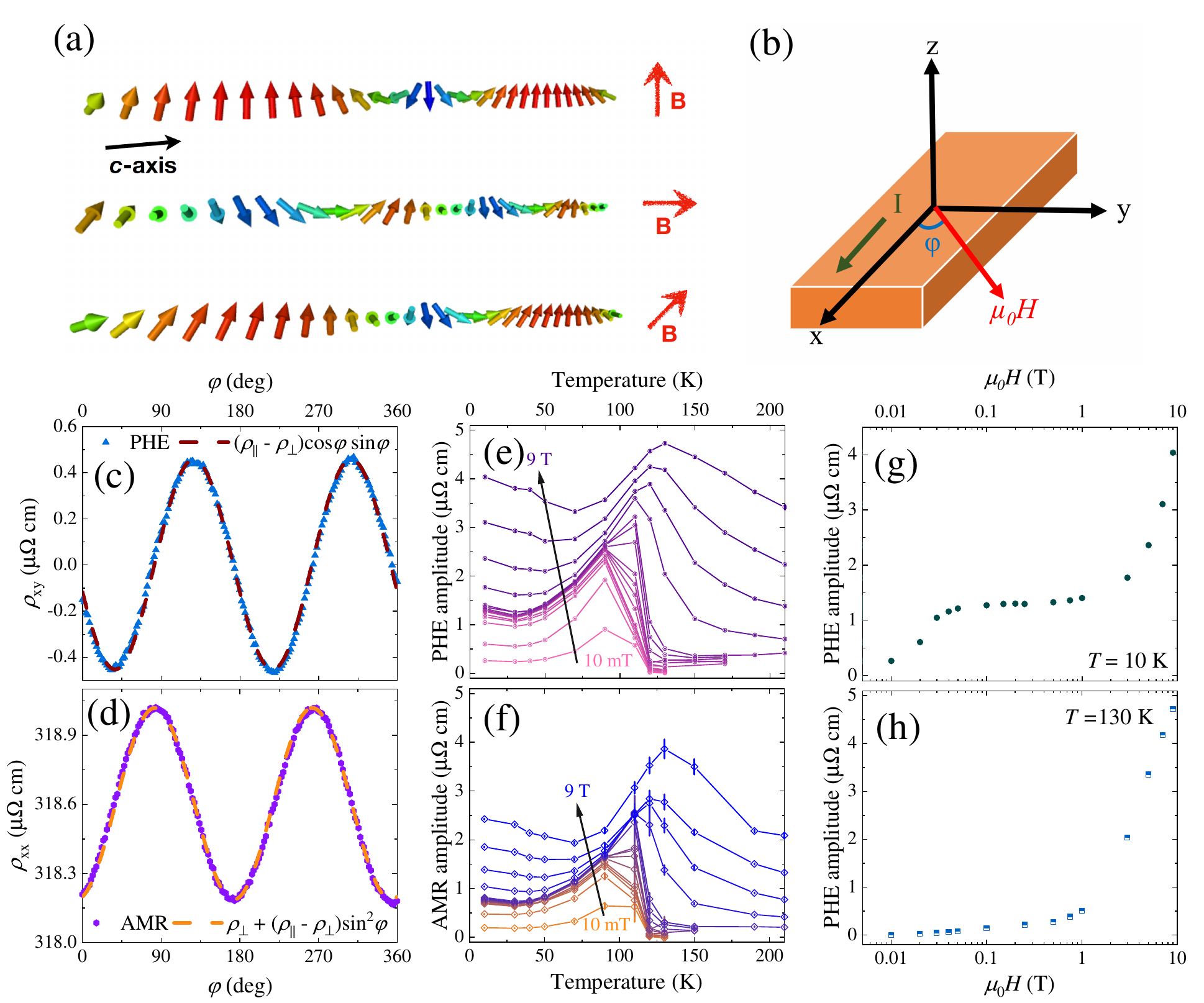}
\caption{(Color online) (a) Possible magnetic ground states as function of the magnetic field direction: CSL, chiral conical phase, and tilted CSL. The chain contains 30~atoms, the model parameters are $J = 10$~meV (exchange), $D = 0.5J$ (DMI),$B = 6$~T with a $45^\circ$ tilt with respect to the $c$~axis. The simulations were performed using the atomistic spin Spirit code~\cite{Gideon:2019}.  Further details can be found in the Appendix.~\ref{APDX: Simulations}.(b) Schematic of the planar Hall effect and anisotropic magnetoresistance set up. (c, d) PHE $\rho_{\mathrm{xy}}$ and AMR $\rho_{\mathrm{xx}}$ taken at 5~K. The dashed lines indicate fits to the PHE and AMR (as described in the text). (e, f) PHE and AMR amplitudes, $\left(\rho_{\parallel}-\rho_{\perp}\right)$, as a function of temperature at various applied fields. The field dependence of the PHE amplitudes at (g) 10 and (h) 130~K.}
\label{FIG: Planar Hall}
\end{figure*}

 One of the keys to understanding the magnetic spin textures in our Hall effect measurements is the presence of a PHE. Indeed, a planar Hall effect and its longitudinal counterpart, an anisotropic magnetoresistance are observed in Cr$_{1/3}$NbS$_{2}$. The presence of the PHE indicates a finite in-plane magnetization in the magnetic structure. Thus, we consider the two possible scenarios currently proposed for Cr$_{1/3}$NbS$_{2}$ when the field is applied close to the $c$~axis: a chiral conical phase (when the offset is less than $1.5^{\circ}$) and a tilted chiral soliton lattice (when the offset is greater than $1.5^{\circ}$)~\cite{Yonemure:2017}. In the CC phase the magnetic moments provide a net magnetic moment that points directly along the helical axis (out-of-plane) due to symmetry. Once the field is sufficient to polarize all the moments along the helical axis the magnetic moments continue to be out-of-plane. Hence, the absence of any in-plane magnetization from the CC phase and eventual FFM phase means no PHE would be observed. In the case of a tilted CSL where all the magnetic moments are uniformly canted towards the helical axis, it is easy to see that there is a net magnetic moment that is neither perpendicular nor parallel to the helical axis. This then means that an in-plane magnetization vector does exist, however, it is tilted slightly out-of-plane. It is important to note that the PHE can persist outside of the tilted CSL phase as the slightly out-of-plane magnetization persists once FFM is stabilized due to any misalignment of the field and the plane. The different magnetic configurations occurring when the external field is applied in different directions are depicted in Fig.~\ref{FIG: Planar Hall}(a). For our conventional Hall effect measurements we apply the field perpendicular to the plane of the platelet, which is assumed to be nominally close to $c$~axis. Taking this symmetry arrangement of current and magnetic field along with the modeled spin structures of the magnetic textures we conclude that the PHE arises from the magnetic moments in the tilted CSL and that they are pointing between the $c$~axis and the $ab$~plane.

For Cr$_{1/3}$NbS$_{2}$ the CSL has been observed via Lorentz transmission electron microscopy (LTEM) when imaging along the $ab$~plane with a field applied perpendicular to the $c$~axis~\cite{Togawa:2013}. For our magnetotransport measurements the current is applied along the length of the platelet, which nominally in the $ab$~plane. For the conventional Hall effect measurements we apply the field perpendicular to the plane of the platelet, which is assumed to be nominally along the $c$~axis. Taking this symmetry arrangement of current and magnetic field along with the modeled spin structures of the magnetic textures described in the previous section we conclude that the magnetic moments in the tilted CSL are pointing between the $c$~axis and the $ab$~plane.

The PHE and AMR as function of angle in Cr$_{1/3}$NbS$_{2}$, is shown in Figs.~\ref{FIG: Planar Hall}(b,c). The angle $\varphi$ between the current and the applied field is defined in Fig.~\ref{FIG: Planar Hall}(b). A clear oscillatory signal of the form $\sin\varphi\cos\varphi$ for the PHE, and $\sin^{2}\varphi$ for the AMR is observed and can be fit by Eq.~\ref{PHE_func}. The amplitude of the oscillations gives the magnitudes of the PHE and AMR. 

The observed temperature and field dependence of the PHE and AMR supports the theory of a tilted-CSL in Cr$_{1/3}$NbS$_{2}$. It can be seen in Figs.~\ref{FIG: Planar Hall}(d,e) that in all fields, close to the magnetic transition, the magnitudes of both the PHE and AMR amplitudes increase. The field dependence of the PHE above and below the magnetic transition are of particular note, as shown in Figs.~\ref{FIG: Planar Hall}(g)~and~\ref{FIG: Planar Hall}(h), as there appear to be two components. Firstly, there exists a low-field component below the magnetic transition, where there is a rapid increase in the PHE amplitude below 100 mT. This component disappears above the transition temperature. We can therefore attribute this to the magnetic state as the transition from the CSL phase to the FFM phase is expected to occur around 100~mT~\cite{Hans:2017}. The high-field component is the traditional component we would expect to observe from a ferromagnetic system.  Indeed, the maximum in the PHE amplitude at fields above 1~T approximately corresponds to the Curie temperature determined from magnetization. To obtain this PHE component  the only requirement is that the sample has some finite in-plane magnetization which we have confirmed in our magnetization measurements (see Fig.~\ref{FIG: Mag and Res}(c). Further planar Hall resistivity as a function of field data across all measured temperatures is shown in the supplemental material~\cite{SMnote}.

A similar oscillatory signal, this time of the form $\sin^{2}\varphi$, is also observed when measuring the longitudinal resistivity in this configuration as shown in Fig.~\ref{FIG: Planar Hall}(d). This longitudinal resistivity is generally referred to as the anisotropic magnetoresistance. From the resistivity tensor it is expected that the amplitude of the PHE and AMR oscillations $\left[\mathrm{i.e.}~\left(\rho_{\parallel}-\rho_{\perp}\right)\right]$ should be of equal magnitude~\cite{Juba:2016}. However, any small misalignment of the sample can cause the ordinary magnetoresistance to leak into the AMR signal. The value for the PHE signal is significantly more reliable, however, as the effects of any leaking of the ordinary Hall effect into the PHE due to a misalignment is antisymmetric and can be removed by averaging the positive and negative field data. \\


\section{Emergence of a giant topological Hall effect}

An anti-symmetrization procedure was first performed on the Hall effect data to remove any symmetric components originating from the magnetoresistance and PHE to simplify the extraction of the THE component. The individual contributions to the Hall signal in Cr$_{1/3}$NbS$_{2}$, namely the OHE and AHE are extracted from the remaining Hall signal following the procedure described in Appendix~\ref{APDX: Hall effect}. A finite and significant component to the Hall signal remains once the OHE and AHE are extracted, we attribute this contribution to the THE. The individual Hall components at several temperatures are shown in Figs.~\ref{FIG: Hall analysis}(a-c) [Further conventional Hall resistivity as a function of field data across all measured temperatures are shown in the supplemental material~\cite{SMnote}.]. The ordinary Hall coefficient $R_{\mathrm{H}}$ is positive across all temperatures indicating hole-like conduction in Cr$_{1/3}$NbS$_{2}$; at 30~K $R_{\mathrm{H}} = 7.31(2)\times10^{-9}$~m$^{3}$/C gives a hole-density of $8.53(2)\times10^{26}$~m$^{-3}$ consistent with what has previously been reported~\cite{Bornstein:2015}. The AHE component is seen to be always negative, diminishing in magnitude with increasing temperature. However, an anomalous Hall component is still noticeable well into the paramagnetic regime. This behavior has previously been observed in MgZnO/ZnO heterostructures and explained using the Giovannini–Kondo model where localized paramagnetic centers can generate an anomalous Hall component from the skew scattering of electrons off these centers~\cite{Maryenko:2017}. The skew, side and intrinsic scattering terms are all found to be significant below the magnetic transition temperature, gradually increasing in magnitude as the temperature decreases. The scattering term can be a useful measure of the quality of crystal used as it is normally only expected to be significant in near perfect crystals~\cite{Onada:2008,Nagaosa:2010review}, hence the high quality of the crystals used in this experiment is particularly notable.

Finally we discuss $\rho_{\mathrm{xy}}^{\mathrm{THE}}$ in the light of CSL for Cr$_{1/3}$NbS$_{2}$. The temperature dependence of the maximum magnitude of $\rho_{\mathrm{xy}}^{\mathrm{THE}}$ is shown in Fig.~\ref{FIG: Hall analysis}(d). At the transition temperature the sign of the THE is seen to flip from positive ($T > T_{\mathrm{c}}$) to negative ($T < T_{\mathrm{c}}$). Below the magnetic transition the topological Hall effect has a maximum magnitude of $2.33~\mu\Omega$~cm at 90~K. The change in the sign of the $\rho_{\mathrm{xy}}^{\mathrm{THE}}$ is likely indicative of a change in the spin polarization of the charge carriers as the temperature passes through the transition. Parallels can be drawn and discussed with the AHE and how it behaves at the phase boundary. It has been shown that due to thermal spin fluctuations, the skew scattering contributions to the AHE become more important and can cause a change of sign~\cite{Ishizuka:2018}. Assuming that the topological Hall effect is similar to the anomalous Hall effect, with the main difference being that the THE emerges from non-coplanar spins instead of spin-orbit coupling, one may conjecture that a similar mechanism occurs at the phase transition. $\rho_{\mathrm{xy}}^{\mathrm{THE}}$ is significantly larger than has been reported in other helical ground state antiferromagnets such as MnSi ($\sim -7$~n$\Omega$~cm)~\cite{Neubauer:2009,Li:2013} and FeGe ($\sim-136$~n$\Omega$~cm)~\cite{Porter:2014}. Both MnSi and FeGe host skyrmion lattices which are thought to only induce a small topological Hall effect. This suggests that the mechanism for the THE in Cr$_{1/3}$NbS$_{2}$ is provided through some other magnetic spin texture, namely the tilted CSL. Similarly, if the rotation period is smaller, then the emergent field is expected to be larger, which can lead to a larger $\rho_{\mathrm{xy}}^{\mathrm{THE}}$. It is expect that the change of period will lead to a non-linear change in the topological charge density once the tilted CSL is stabilized since $\mathcal{C}_{ijk}=\vec{m}_{i}\cdot(\vec{m}_{j}\times\vec{m}_{k})$, hence important changes in the THE will occur. Furthermore, the topological Hall effect is a complex phenomena and depends on other details of the electronic structure such as the scattering rate off the magnetic atoms and the density of the charge carriers. A contour map of $\rho_{\mathrm{xy}}^{\mathrm{THE}}$ [see Fig.~\ref{FIG: Hall analysis}(e)] reveals the presence of $\rho_{\mathrm{xy}}^{\mathrm{THE}}$ across most temperatures below the transition temperature before becoming negligible at 2~K. As the temperature decreases, the fluctuations in the magnetic spins from the chiral phase would be expected to reduce. Therefore, as the temperature is reduced the magnitude of the THE would be expected to reduce as well. 
The region of increased magnitude in $\rho_{\mathrm{xy}}^{\mathrm{THE}}$ corresponds to the same region where the CSL has been reported by LTEM~\cite{Togawa:2012,Togawa:2015,Paterson:2020} and ac magnetization~\cite{Tsuruta:2016}. This strongly suggests that the $\rho_{\mathrm{xy}}^{\mathrm{THE}}$ is due to a non-coplanar spin structure arising from a tilted CSL. \\

\section{Summary}

Remarkably, the chiral helimagnet Cr$_{1/3}$NbS$_{2}$ is found to host both a giant THE and PHE. We show that the observation of a PHE and THE in Cr$_{1/3}$NbS$_{2}$ indicates the presence of a tilted CSL. Furthermore, the close correspondence of the PHE and AMR in this complex non-coplanar magnet validates the theory that the connection between these two quantities still holds in these more complex systems. The conventional Hall measurements performed in this study demonstrate that non-coplanar spins arising from a tilted CSL can indeed generate the emergent field necessary for the THE. Studying Cr$_{1/3}$NbS$_{2}$, where the phase diagram has been extensively mapped out, has allowed a one-to-one correspondence to be drawn between the relative magnetic phases and magneto-transport properties observed. Notably, the THE exhibited in Cr$_{1/3}$NbS$_{2}$ is several orders of magnitude larger than those of similar helimagnetic materials where skyrmions are observed. This study marks an important step forward in demonstrating that magneto-transport measurements, and more specifically the THE, can be used to detect not just skyrmions but also non-coplanar spin textures such as a tilted CSL.
\\
Data will be made available via Warwick Research Archive Portal~\cite{data_portal}.

\appendix

\section{Fitting procedure for the Hall effect} \label{APDX: Hall effect}

To estimate the contribution the topological Hall effect makes to the Hall resistivity, the following fitting procedure was used. To simplify the problem we first remove any Hall contributions from the PHE and any magnetoresistance. To do this an antisymmetrization procedure was performed using $\rho(+H)' = [\rho(+H^{\mathrm{inc}})-\rho(-H^{\mathrm{dec}})]/2$ and $\rho(-H)' = [\rho(-H^{\mathrm{inc}})-\rho(+H^{\mathrm{dec}})]/2$ where $\rho(+H^{\mathrm{inc}})$ is the Hall resistivity in the positive field quadrants when the field is increasing in magnitude, $\rho(+H^{\mathrm{dec}})$ is the Hall resistivity in the positive field quadrants when the field is decreasing in magnitude, $\rho(-H^{\mathrm{inc}})$ is the Hall resistivity in the negative field quadrants when the field is increasing in magnitude and $\rho(-H^{\mathrm{inc}})$ is the Hall resistivity in the negative field quadrants when the field is decreasing in magnitude. For fields greater than some critical field value, $\rho_{\mathrm{xy}}^{\mathrm{THE}}$ should be equal to zero as forced ferromagnetism should have no spin chirality. The contribution from the ordinary Hall effect $\rho_{\mathrm{xy}}^{\mathrm{OHE}}$ and the anomalous Hall effect $\rho_{\mathrm{xy}}^{\mathrm{AHE}}$ can be calculated by linearizing the data using $\rho_{\mathrm{xy}}/B = \alpha\rho_{\mathrm{xx}}^{2}M/H + R_{\mathrm{H}}$ (not shown). Here $\alpha = \left(S_{\mathrm{Int}} + S_{\mathrm{Side}}\right)$, where $S_{\mathrm{Int}}$ is the intrinsic contribution and $S_{\mathrm{Side}}$ is the side jump contribution, only provides the $\rho_{\mathrm{xx}}^{2}$ contribution for the AHE but we now have a value for $R_{\mathrm{H}}$. To estimate the full contribution of $\rho_{\mathrm{xy}}^{\mathrm{AHE}}$ we can subtract $\rho_{\mathrm{xy}}^{\mathrm{OHE}}$ from the total Hall resistivity and by using the following equality $\left(\rho_{\mathrm{xy}}-\rho_{xy}^{\mathrm{OHE}}\right)/\rho_{\mathrm{xx}}M = \alpha\rho_{\mathrm{xx}} + S_{\mathrm{Skew}}$, where $S_{\mathrm{Skew}}$ is the skew scattering contribution, the data can be linearized. This allows for $\alpha$ and $S_{\mathrm{Skew}}$ to be extracted, (not shown).

\section{Simulations of the Spin Dynamics} \label{APDX: Simulations}

The magnetic state of the system is described by an extended Heisenberg Hamiltonian:\\
\begin{equation}
\mathcal{E}_{M} = -\frac{1}{2}\sum_{ij}J_{ij}\,\vec{M}_{i}\cdot\vec{M}_{j} + \frac{1}{2}\sum_{ij}\vec{D}_{ij}\cdot(\vec{M}_i\times\vec{M}_j)- \sum_{i}\vec{M}_{i}\cdot\vec{B}\quad. 
\label{Magnetic_energ}
\end{equation}
The first term in the Hamiltonian designates the isotropic Heisenberg exchange interaction, the second is the Dzyaloshinskii-Moriya interaction, and the third is the Zeeman coupling to an external magnetic field. The following parameters are used: a nearest neighbor exchange $J_{ij}=10.0$ meV, and nearest neighbor DMI $D=5.0$ meV, and magnetic moment of $M= 3~\mu_{\mathrm{B}}$ with magnetic field of $6$\,T. The field is oriented in two directions in practice either along the $z$ or tilted at $45^\circ$ away from the $z$-axis in the $xz$ plane. The minimization of the energy is performed using the Spirit code, which uses an algorithm based on the Landau-Lifshitz-Gilbert equation~\cite{Gideon:2019}.

\section{Topological Hall Effect from Chiral Phases} \label{APDX: THE}

\begin{figure}[h]
    \centering
	\includegraphics[width=0.65\columnwidth]{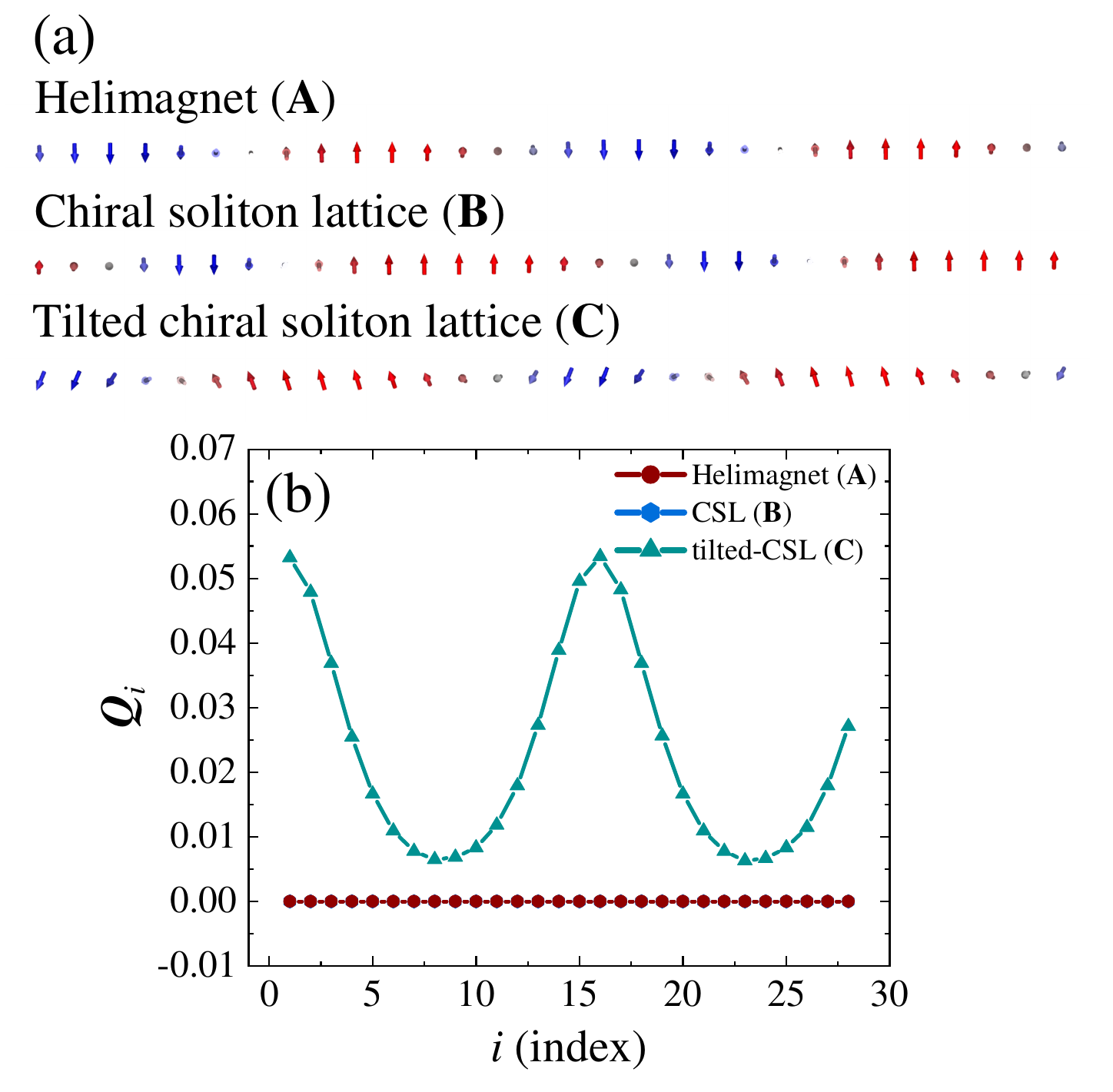}
	\caption{(a) Three spin configurations possible in Cr$_{1/3}$NbS$_{2}$: Helimagnet (\textbf{A}), chiral soliton lattice (\textbf{B}), and tilted chiral soliton lattice (\textbf{C}). (b) Evolution of the topological charge density for a Helimagnet (\textbf{A}), a chiral soliton lattice (\textbf{B}) and a tilted chiral soliton lattice (\textbf{C}). The topological charge density for the helimagnet state (\textbf{A}) is zero for all \textit{i} so is not visible.}
	\label{FIG: Topological Charge Density}
\end{figure}

As discussed in the previous section, neither the helimagnetic phase nor the chiral soliton lattice will display a topological Hall effect since the scalar chirality must vanish, \textit{i.e.} $\mathcal{C}_{ijk}= \vec{m}_{i}\cdot(\vec{m}_{j}\times\vec{m}_{k})$. However, for the case of a tilted-CSL, a finite topological Hall effect contribution is possible. To illustrate this we consider three different cases: helimagnet (\textbf{A}), CSL (\textbf{B}) and a tilted-CSL (\textbf{C}), these magnetic states are shown in Fig.~\ref{FIG: Topological Charge Density}(a). Using these configurations, we evaluate  the topological charge density between nearest neighbors:
\begin{equation}
\mathbf{Q}_{i} = \vec{m}_{i-1}\cdot\left(\vec{m}_{i}\times\vec{m}_{i+1}\right). 
\end{equation}
The results are shown in Fig.~\ref{FIG: Topological Charge Density}(b), where we found that $\mathbf{Q}_{i}$ systematically vanishes for the configurations (\textbf{A}) and (\textbf{B}), while a finite contribution is present for (\textbf{C}).  Thus, as a consequence of applying a magnetic field in the $(xz)$ plane the magnetic state becomes non-coplanar. Furthermore, in these metallic systems, coupling beyond nearest neighbors is also possible, thus more contributions to the THE are present. To summarize, both the PHE and THE are consequences of the tilted CSL. Furthermore, a vector chirality contribution $(\vec{m}_i\times\vec{m}_j)$ to the Hall effect can be relevant in the non-centrosymmetric systems treated in this manuscript, this was discussed in Ref.~\cite{Bouaziz:2021}. However, for the sake of simplicity, we only interpret the experimental data in terms of the THE.

\begin{acknowledgments}
We would like to acknowledge Tom Orton, Patrick Ruddy and Daisy Ashworth for their technical support. This work was financially supported by three Engineering and Physical Sciences Research Council grants: EP/T005963/1, EP/M028941/1 (PRETAMAG), and the UK Skyrmion Project Grant, EP/N032128/1. 
\end{acknowledgments}
 
\bibliography{CrNb3S6_References}

\begin{thebibliography}{62}%
\makeatletter
\providecommand \@ifxundefined [1]{%
 \@ifx{#1\undefined}
}%
\providecommand \@ifnum [1]{%
 \ifnum #1\expandafter \@firstoftwo
 \else \expandafter \@secondoftwo
 \fi
}%
\providecommand \@ifx [1]{%
 \ifx #1\expandafter \@firstoftwo
 \else \expandafter \@secondoftwo
 \fi
}%
\providecommand \natexlab [1]{#1}%
\providecommand \enquote  [1]{``#1''}%
\providecommand \bibnamefont  [1]{#1}%
\providecommand \bibfnamefont [1]{#1}%
\providecommand \citenamefont [1]{#1}%
\providecommand \href@noop [0]{\@secondoftwo}%
\providecommand \href [0]{\begingroup \@sanitize@url \@href}%
\providecommand \@href[1]{\@@startlink{#1}\@@href}%
\providecommand \@@href[1]{\endgroup#1\@@endlink}%
\providecommand \@sanitize@url [0]{\catcode `\\12\catcode `\$12\catcode
  `\&12\catcode `\#12\catcode `\^12\catcode `\_12\catcode `\%12\relax}%
\providecommand \@@startlink[1]{}%
\providecommand \@@endlink[0]{}%
\providecommand \url  [0]{\begingroup\@sanitize@url \@url }%
\providecommand \@url [1]{\endgroup\@href {#1}{\urlprefix }}%
\providecommand \urlprefix  [0]{URL }%
\providecommand \Eprint [0]{\href }%
\providecommand \doibase [0]{https://doi.org/}%
\providecommand \selectlanguage [0]{\@gobble}%
\providecommand \bibinfo  [0]{\@secondoftwo}%
\providecommand \bibfield  [0]{\@secondoftwo}%
\providecommand \translation [1]{[#1]}%
\providecommand \BibitemOpen [0]{}%
\providecommand \bibitemStop [0]{}%
\providecommand \bibitemNoStop [0]{.\EOS\space}%
\providecommand \EOS [0]{\spacefactor3000\relax}%
\providecommand \BibitemShut  [1]{\csname bibitem#1\endcsname}%
\let\auto@bib@innerbib\@empty
\bibitem [{\citenamefont {Ningrum}\ \emph {et~al.}(2020)\citenamefont
  {Ningrum}, \citenamefont {Liu}, \citenamefont {Wang}, \citenamefont {Yin},
  \citenamefont {CAo}, \citenamefont {Zha}, \citenamefont {Xie}, \citenamefont
  {HJiang}, \citenamefont {Sun}, \citenamefont {Qin}, \citenamefont {Chen},
  \citenamefont {Qin}, \citenamefont {Zhu}, \citenamefont {Qang},\ and\
  \citenamefont {Huang}}]{Ningrum:2020}%
  \BibitemOpen
  \bibfield  {author} {\bibinfo {author} {\bibfnamefont {V.~P.}\ \bibnamefont
  {Ningrum}}, \bibinfo {author} {\bibfnamefont {B.}~\bibnamefont {Liu}},
  \bibinfo {author} {\bibfnamefont {W.}~\bibnamefont {Wang}}, \bibinfo {author}
  {\bibfnamefont {Y.}~\bibnamefont {Yin}}, \bibinfo {author} {\bibfnamefont
  {Y.}~\bibnamefont {CAo}}, \bibinfo {author} {\bibfnamefont {C.}~\bibnamefont
  {Zha}}, \bibinfo {author} {\bibfnamefont {H.}~\bibnamefont {Xie}}, \bibinfo
  {author} {\bibfnamefont {X.}~\bibnamefont {HJiang}}, \bibinfo {author}
  {\bibfnamefont {Y.}~\bibnamefont {Sun}}, \bibinfo {author} {\bibfnamefont
  {S.}~\bibnamefont {Qin}}, \bibinfo {author} {\bibfnamefont {X.}~\bibnamefont
  {Chen}}, \bibinfo {author} {\bibfnamefont {T.}~\bibnamefont {Qin}}, \bibinfo
  {author} {\bibfnamefont {C.}~\bibnamefont {Zhu}}, \bibinfo {author}
  {\bibfnamefont {L.}~\bibnamefont {Qang}},\ and\ \bibinfo {author}
  {\bibfnamefont {W.}~\bibnamefont {Huang}},\ }\bibfield  {title} {\bibinfo
  {title} {Recent advances in two-dimensional magnets: Physics and devices
  towards spintronic applications.},\ }\href
  {https://doi.org/10.34133/2020/1768918} {\bibfield  {journal} {\bibinfo
  {journal} {AAAS Research}\ }\textbf {\bibinfo {volume} {2020}},\ \bibinfo
  {pages} {1768918} (\bibinfo {year} {2020})}\BibitemShut {NoStop}%
\bibitem [{\citenamefont {Mak}\ \emph {et~al.}(2019)\citenamefont {Mak},
  \citenamefont {Shan},\ and\ \citenamefont {Ralph}}]{Mak:2019}%
  \BibitemOpen
  \bibfield  {author} {\bibinfo {author} {\bibfnamefont {K.~F.}\ \bibnamefont
  {Mak}}, \bibinfo {author} {\bibfnamefont {J.}~\bibnamefont {Shan}},\ and\
  \bibinfo {author} {\bibfnamefont {D.~C.}\ \bibnamefont {Ralph}},\ }\bibfield
  {title} {\bibinfo {title} {Probing and controlling magnetic states in {2D}
  layered magnetic materials.},\ }\href
  {https://doi.org/10.1038/s42254-019-0110-y} {\bibfield  {journal} {\bibinfo
  {journal} {Nat. Rev. Phys.}\ }\textbf {\bibinfo {volume} {1}},\ \bibinfo
  {pages} {646} (\bibinfo {year} {2019})}\BibitemShut {NoStop}%
\bibitem [{\citenamefont {M\"{u}hlbauer}\ \emph {et~al.}(2009)\citenamefont
  {M\"{u}hlbauer}, \citenamefont {Binz}, \citenamefont {Jonietz}, \citenamefont
  {Pfleiderer}, \citenamefont {Rosch}, \citenamefont {Neubauer}, \citenamefont
  {Georgii},\ and\ \citenamefont {B\"{o}ni}}]{Muhlbauer:2009}%
  \BibitemOpen
  \bibfield  {author} {\bibinfo {author} {\bibfnamefont {S.}~\bibnamefont
  {M\"{u}hlbauer}}, \bibinfo {author} {\bibfnamefont {B.}~\bibnamefont {Binz}},
  \bibinfo {author} {\bibfnamefont {F.}~\bibnamefont {Jonietz}}, \bibinfo
  {author} {\bibfnamefont {C.}~\bibnamefont {Pfleiderer}}, \bibinfo {author}
  {\bibfnamefont {A.}~\bibnamefont {Rosch}}, \bibinfo {author} {\bibfnamefont
  {A.}~\bibnamefont {Neubauer}}, \bibinfo {author} {\bibfnamefont
  {R.}~\bibnamefont {Georgii}},\ and\ \bibinfo {author} {\bibfnamefont
  {P.}~\bibnamefont {B\"{o}ni}},\ }\bibfield  {title} {\bibinfo {title}
  {Skyrmion lattice in a chiral magnet.},\ }\href
  {https://doi.org/10.1126/science.1166767} {\bibfield  {journal} {\bibinfo
  {journal} {Science}\ }\textbf {\bibinfo {volume} {323}},\ \bibinfo {pages}
  {915} (\bibinfo {year} {2009})}\BibitemShut {NoStop}%
\bibitem [{\citenamefont {Karube}\ \emph {et~al.}(2016)\citenamefont {Karube},
  \citenamefont {White}, \citenamefont {Reynolds}, \citenamefont {Gavilano},
  \citenamefont {Oike}, \citenamefont {Kikkawa}, \citenamefont {Kagawa},
  \citenamefont {Tokunaga}, \citenamefont {Ronnow}, \citenamefont {Tokura},\
  and\ \citenamefont {Taguchi}}]{Karube:2016}%
  \BibitemOpen
  \bibfield  {author} {\bibinfo {author} {\bibfnamefont {K.}~\bibnamefont
  {Karube}}, \bibinfo {author} {\bibfnamefont {J.~S.}\ \bibnamefont {White}},
  \bibinfo {author} {\bibfnamefont {N.}~\bibnamefont {Reynolds}}, \bibinfo
  {author} {\bibfnamefont {J.~L.}\ \bibnamefont {Gavilano}}, \bibinfo {author}
  {\bibfnamefont {H.}~\bibnamefont {Oike}}, \bibinfo {author} {\bibfnamefont
  {A.}~\bibnamefont {Kikkawa}}, \bibinfo {author} {\bibfnamefont
  {F.}~\bibnamefont {Kagawa}}, \bibinfo {author} {\bibfnamefont
  {Y.}~\bibnamefont {Tokunaga}}, \bibinfo {author} {\bibfnamefont {H.~M.}\
  \bibnamefont {Ronnow}}, \bibinfo {author} {\bibfnamefont {Y.}~\bibnamefont
  {Tokura}},\ and\ \bibinfo {author} {\bibfnamefont {Y.}~\bibnamefont
  {Taguchi}},\ }\bibfield  {title} {\bibinfo {title} {Robust metastable
  skyrmions and their triangular–square lattice structural transition in a
  high-temperature chiral magnet.},\ }\href
  {https://www.nature.com/articles/nmat4752} {\bibfield  {journal} {\bibinfo
  {journal} {Nat. Mater.}\ }\textbf {\bibinfo {volume} {15}},\ \bibinfo {pages}
  {1237} (\bibinfo {year} {2016})}\BibitemShut {NoStop}%
\bibitem [{\citenamefont {K\'{e}zsmarki}\ \emph {et~al.}(2015)\citenamefont
  {K\'{e}zsmarki}, \citenamefont {Bord\'{a}cs}, \citenamefont {Milde},
  \citenamefont {Neuber}, \citenamefont {Eng}, \citenamefont {White},
  \citenamefont {R\o{}nnow}, \citenamefont {Dewhurst}, \citenamefont
  {Mochizuki}, \citenamefont {Yanai}, \citenamefont {Nakamura}, \citenamefont
  {Ehlers}, \citenamefont {Tsurkan},\ and\ \citenamefont
  {Loidl}}]{Kezsmarki:2015}%
  \BibitemOpen
  \bibfield  {author} {\bibinfo {author} {\bibfnamefont {I.}~\bibnamefont
  {K\'{e}zsmarki}}, \bibinfo {author} {\bibfnamefont {S.}~\bibnamefont
  {Bord\'{a}cs}}, \bibinfo {author} {\bibfnamefont {P.}~\bibnamefont {Milde}},
  \bibinfo {author} {\bibfnamefont {E.}~\bibnamefont {Neuber}}, \bibinfo
  {author} {\bibfnamefont {L.~M.}\ \bibnamefont {Eng}}, \bibinfo {author}
  {\bibfnamefont {J.~S.}\ \bibnamefont {White}}, \bibinfo {author}
  {\bibfnamefont {H.~M.}\ \bibnamefont {R\o{}nnow}}, \bibinfo {author}
  {\bibfnamefont {C.~D.}\ \bibnamefont {Dewhurst}}, \bibinfo {author}
  {\bibfnamefont {M.}~\bibnamefont {Mochizuki}}, \bibinfo {author}
  {\bibfnamefont {K.}~\bibnamefont {Yanai}}, \bibinfo {author} {\bibfnamefont
  {H.}~\bibnamefont {Nakamura}}, \bibinfo {author} {\bibfnamefont
  {D.}~\bibnamefont {Ehlers}}, \bibinfo {author} {\bibfnamefont
  {V.}~\bibnamefont {Tsurkan}},\ and\ \bibinfo {author} {\bibfnamefont
  {A.}~\bibnamefont {Loidl}},\ }\bibfield  {title} {\bibinfo {title}
  {{Néel}-type skyrmion lattice with confined orientation in the polar
  magnetic semiconductor {GaV$_{4}$S$_{8}$}.},\ }\href
  {https://doi.org/10.1038/nmat4402} {\bibfield  {journal} {\bibinfo  {journal}
  {Nat. Mat.}\ }\textbf {\bibinfo {volume} {14}},\ \bibinfo {pages} {1116}
  (\bibinfo {year} {2015})}\BibitemShut {NoStop}%
\bibitem [{\citenamefont {Saha}\ \emph {et~al.}(2019)\citenamefont {Saha},
  \citenamefont {Srivastava}, \citenamefont {Ma}, \citenamefont {Jena},
  \citenamefont {Werner}, \citenamefont {Kumar}, \citenamefont {Felser},\ and\
  \citenamefont {Parkin}}]{Saha:2019}%
  \BibitemOpen
  \bibfield  {author} {\bibinfo {author} {\bibfnamefont {R.}~\bibnamefont
  {Saha}}, \bibinfo {author} {\bibfnamefont {A.~K.}\ \bibnamefont
  {Srivastava}}, \bibinfo {author} {\bibfnamefont {T.}~\bibnamefont {Ma}},
  \bibinfo {author} {\bibfnamefont {J.}~\bibnamefont {Jena}}, \bibinfo {author}
  {\bibfnamefont {P.}~\bibnamefont {Werner}}, \bibinfo {author} {\bibfnamefont
  {V.}~\bibnamefont {Kumar}}, \bibinfo {author} {\bibfnamefont
  {C.}~\bibnamefont {Felser}},\ and\ \bibinfo {author} {\bibfnamefont
  {S.~S.~P.}\ \bibnamefont {Parkin}},\ }\bibfield  {title} {\bibinfo {title}
  {Intrinsic stability of magnetic anti-skyrmions in the tetragonal inverse
  {Heusler} compound {Mn$_{1.4}$Pt$_{0.9}$Pd$_{0.1}$Sn}.},\ }\href
  {https://www.nature.com/articles/s41467-019-13323-x} {\bibfield  {journal}
  {\bibinfo  {journal} {Nat. Comm.}\ }\textbf {\bibinfo {volume} {10}},\
  \bibinfo {pages} {5305} (\bibinfo {year} {2019})}\BibitemShut {NoStop}%
\bibitem [{\citenamefont {Lui}\ \emph {et~al.}(2016)\citenamefont {Lui},
  \citenamefont {Puliafito}, \citenamefont {Montaigne}, \citenamefont {Petit},
  \citenamefont {Deranlot}, \citenamefont {Andrieu}, \citenamefont {Ozatay},
  \citenamefont {Finocchio},\ and\ \citenamefont {Hauet}}]{Liu:2016}%
  \BibitemOpen
  \bibfield  {author} {\bibinfo {author} {\bibfnamefont {T.}~\bibnamefont
  {Lui}}, \bibinfo {author} {\bibfnamefont {V.}~\bibnamefont {Puliafito}},
  \bibinfo {author} {\bibfnamefont {F.}~\bibnamefont {Montaigne}}, \bibinfo
  {author} {\bibfnamefont {S.}~\bibnamefont {Petit}}, \bibinfo {author}
  {\bibfnamefont {C.}~\bibnamefont {Deranlot}}, \bibinfo {author}
  {\bibfnamefont {S.}~\bibnamefont {Andrieu}}, \bibinfo {author} {\bibfnamefont
  {O.}~\bibnamefont {Ozatay}}, \bibinfo {author} {\bibfnamefont
  {G.}~\bibnamefont {Finocchio}},\ and\ \bibinfo {author} {\bibfnamefont
  {T.}~\bibnamefont {Hauet}},\ }\bibfield  {title} {\bibinfo {title}
  {Reproducible formation of single magnetic bubbles in an array of patterned
  dots.},\ }\href {https://doi.org/10.1088/0022-3727/49/24/245002} {\bibfield
  {journal} {\bibinfo  {journal} {J. Phys D: Appl. Phys.}\ }\textbf {\bibinfo
  {volume} {49}},\ \bibinfo {pages} {245002} (\bibinfo {year}
  {2016})}\BibitemShut {NoStop}%
\bibitem [{\citenamefont {Pribaig}\ \emph {et~al.}(2007)\citenamefont
  {Pribaig}, \citenamefont {Krivorotov}, \citenamefont {Fuchs}, \citenamefont
  {Braganca}, \citenamefont {Ozatay}, \citenamefont {Sankey}, \citenamefont
  {Ralph},\ and\ \citenamefont {Buhrman}}]{Pribaig:2007}%
  \BibitemOpen
  \bibfield  {author} {\bibinfo {author} {\bibfnamefont {V.~S.}\ \bibnamefont
  {Pribaig}}, \bibinfo {author} {\bibfnamefont {I.~N.}\ \bibnamefont
  {Krivorotov}}, \bibinfo {author} {\bibfnamefont {G.~D.}\ \bibnamefont
  {Fuchs}}, \bibinfo {author} {\bibfnamefont {P.~M.}\ \bibnamefont {Braganca}},
  \bibinfo {author} {\bibfnamefont {O.}~\bibnamefont {Ozatay}}, \bibinfo
  {author} {\bibfnamefont {J.~C.}\ \bibnamefont {Sankey}}, \bibinfo {author}
  {\bibfnamefont {D.~C.}\ \bibnamefont {Ralph}},\ and\ \bibinfo {author}
  {\bibfnamefont {R.~A.}\ \bibnamefont {Buhrman}},\ }\bibfield  {title}
  {\bibinfo {title} {Magnetic vortex oscillator driven by d.c. spin-polarized
  current.},\ }\href {https://doi.org/10.1038/nphys619} {\bibfield  {journal}
  {\bibinfo  {journal} {Nat. Phys.}\ }\textbf {\bibinfo {volume} {3}},\
  \bibinfo {pages} {498} (\bibinfo {year} {2007})}\BibitemShut {NoStop}%
\bibitem [{\citenamefont {Dzyaloshinsky}(1958)}]{Dzyaloshinky:1958}%
  \BibitemOpen
  \bibfield  {author} {\bibinfo {author} {\bibfnamefont {I.}~\bibnamefont
  {Dzyaloshinsky}},\ }\bibfield  {title} {\bibinfo {title} {A thermodynamic
  theory of “weak” ferromagnetism of antiferromagnetics.},\ }\href
  {https://doi.org/10.1016/0022-3697(58)90076-3} {\bibfield  {journal}
  {\bibinfo  {journal} {J. Phys. Chem. Solids}\ }\textbf {\bibinfo {volume}
  {4}},\ \bibinfo {pages} {241} (\bibinfo {year} {1958})}\BibitemShut {NoStop}%
\bibitem [{\citenamefont {Moriya}(1960)}]{Moriya:1960}%
  \BibitemOpen
  \bibfield  {author} {\bibinfo {author} {\bibfnamefont {T.}~\bibnamefont
  {Moriya}},\ }\bibfield  {title} {\bibinfo {title} {Anisotropic superexchange
  interaction and weak ferromagnetism.},\ }\href
  {https://doi.org/10.1103/PhysRev.120.91} {\bibfield  {journal} {\bibinfo
  {journal} {Phys. Rev.}\ }\textbf {\bibinfo {volume} {120}},\ \bibinfo {pages}
  {91} (\bibinfo {year} {1960})}\BibitemShut {NoStop}%
\bibitem [{\citenamefont {Heinze}\ \emph {et~al.}(2011)\citenamefont {Heinze},
  \citenamefont {von Bergmann}, \citenamefont {Menzel}, \citenamefont {Brede},
  \citenamefont {Kubetzka}, \citenamefont {Wiesendanger}, \citenamefont
  {Bihlmayer},\ and\ \citenamefont {Bl{\"u}gel}}]{Heinze:2011spontaneous}%
  \BibitemOpen
  \bibfield  {author} {\bibinfo {author} {\bibfnamefont {S.}~\bibnamefont
  {Heinze}}, \bibinfo {author} {\bibfnamefont {K.}~\bibnamefont {von
  Bergmann}}, \bibinfo {author} {\bibfnamefont {M.}~\bibnamefont {Menzel}},
  \bibinfo {author} {\bibfnamefont {J.}~\bibnamefont {Brede}}, \bibinfo
  {author} {\bibfnamefont {A.}~\bibnamefont {Kubetzka}}, \bibinfo {author}
  {\bibfnamefont {R.}~\bibnamefont {Wiesendanger}}, \bibinfo {author}
  {\bibfnamefont {G.}~\bibnamefont {Bihlmayer}},\ and\ \bibinfo {author}
  {\bibfnamefont {S.}~\bibnamefont {Bl{\"u}gel}},\ }\bibfield  {title}
  {\bibinfo {title} {Spontaneous atomic-scale magnetic skyrmion lattice in two
  dimensions.},\ }\href {https://doi.org/10.1038/nphys2045} {\bibfield
  {journal} {\bibinfo  {journal} {Nat. Phys.}\ }\textbf {\bibinfo {volume}
  {7}},\ \bibinfo {pages} {713} (\bibinfo {year} {2011})}\BibitemShut {NoStop}%
\bibitem [{\citenamefont {Ozawa}\ \emph {et~al.}(2017)\citenamefont {Ozawa},
  \citenamefont {Hayami},\ and\ \citenamefont {Motome}}]{Ozawa:2017}%
  \BibitemOpen
  \bibfield  {author} {\bibinfo {author} {\bibfnamefont {R.}~\bibnamefont
  {Ozawa}}, \bibinfo {author} {\bibfnamefont {S.}~\bibnamefont {Hayami}},\ and\
  \bibinfo {author} {\bibfnamefont {Y.}~\bibnamefont {Motome}},\ }\bibfield
  {title} {\bibinfo {title} {Zero-field skyrmions with a high topological
  number in itinerant magnets.},\ }\href
  {https://doi.org/10.1103/PhysRevLett.118.147205} {\bibfield  {journal}
  {\bibinfo  {journal} {Phys. Rev. Lett.}\ }\textbf {\bibinfo {volume} {118}},\
  \bibinfo {pages} {147205} (\bibinfo {year} {2017})}\BibitemShut {NoStop}%
\bibitem [{\citenamefont {Grytsiuk}\ \emph {et~al.}(2020)\citenamefont
  {Grytsiuk}, \citenamefont {Hanke}, \citenamefont {Hoffmann}, \citenamefont
  {Bouaziz}, \citenamefont {Gomonay}, \citenamefont {Bihlmayer}, \citenamefont
  {Lounis}, \citenamefont {Mokrousov},\ and\ \citenamefont
  {Bl{\"u}gel}}]{grytsiuk2020topological}%
  \BibitemOpen
  \bibfield  {author} {\bibinfo {author} {\bibfnamefont {S.}~\bibnamefont
  {Grytsiuk}}, \bibinfo {author} {\bibfnamefont {J.-P.}\ \bibnamefont {Hanke}},
  \bibinfo {author} {\bibfnamefont {M.}~\bibnamefont {Hoffmann}}, \bibinfo
  {author} {\bibfnamefont {J.}~\bibnamefont {Bouaziz}}, \bibinfo {author}
  {\bibfnamefont {O.}~\bibnamefont {Gomonay}}, \bibinfo {author} {\bibfnamefont
  {G.}~\bibnamefont {Bihlmayer}}, \bibinfo {author} {\bibfnamefont
  {S.}~\bibnamefont {Lounis}}, \bibinfo {author} {\bibfnamefont
  {Y.}~\bibnamefont {Mokrousov}},\ and\ \bibinfo {author} {\bibfnamefont
  {S.}~\bibnamefont {Bl{\"u}gel}},\ }\bibfield  {title} {\bibinfo {title}
  {Topological-chiral magnetic interactions driven by emergent orbital
  magnetism},\ }\href {https://doi.org/10.1038/s41467-019-14030-3} {\bibfield
  {journal} {\bibinfo  {journal} {Nat. Commun.}\ }\textbf {\bibinfo {volume}
  {11}},\ \bibinfo {pages} {511} (\bibinfo {year} {2020})}\BibitemShut
  {NoStop}%
\bibitem [{\citenamefont {Okubo}\ \emph {et~al.}(2012)\citenamefont {Okubo},
  \citenamefont {Chung},\ and\ \citenamefont {Kawamura}}]{Okubo:2012}%
  \BibitemOpen
  \bibfield  {author} {\bibinfo {author} {\bibfnamefont {T.}~\bibnamefont
  {Okubo}}, \bibinfo {author} {\bibfnamefont {S.}~\bibnamefont {Chung}},\ and\
  \bibinfo {author} {\bibfnamefont {H.}~\bibnamefont {Kawamura}},\ }\bibfield
  {title} {\bibinfo {title} {Multiple-$q$ states and the skyrmion lattice of
  the triangular-lattice {Heisenberg} antiferromagnet under magnetic fields.},\
  }\href {https://doi.org/10.1103/PhysRevLett.108.017206} {\bibfield  {journal}
  {\bibinfo  {journal} {Phys. Rev. Lett.}\ }\textbf {\bibinfo {volume} {108}},\
  \bibinfo {pages} {017206} (\bibinfo {year} {2012})}\BibitemShut {NoStop}%
\bibitem [{\citenamefont {Leonov}\ and\ \citenamefont
  {Mostovoy}(2015)}]{Leonov:2015multiply}%
  \BibitemOpen
  \bibfield  {author} {\bibinfo {author} {\bibfnamefont {A.~O.}\ \bibnamefont
  {Leonov}}\ and\ \bibinfo {author} {\bibfnamefont {M.}~\bibnamefont
  {Mostovoy}},\ }\bibfield  {title} {\bibinfo {title} {Multiply periodic states
  and isolated skyrmions in an anisotropic frustrated magnet.},\ }\href
  {https://doi.org/10.1038/ncomms9275} {\bibfield  {journal} {\bibinfo
  {journal} {Nat. Comm.}\ }\textbf {\bibinfo {volume} {6}},\ \bibinfo {pages}
  {8275} (\bibinfo {year} {2015})}\BibitemShut {NoStop}%
\bibitem [{\citenamefont {Kanazawa}\ \emph {et~al.}(2011)\citenamefont
  {Kanazawa}, \citenamefont {Onose}, \citenamefont {Arima}, \citenamefont
  {Okuyama}, \citenamefont {Ohoyama}, \citenamefont {Wakimoto}, \citenamefont
  {Kakurai}, \citenamefont {Ishiwata},\ and\ \citenamefont
  {Tokura}}]{Kanazawa:2011}%
  \BibitemOpen
  \bibfield  {author} {\bibinfo {author} {\bibfnamefont {N.}~\bibnamefont
  {Kanazawa}}, \bibinfo {author} {\bibfnamefont {Y.}~\bibnamefont {Onose}},
  \bibinfo {author} {\bibfnamefont {T.}~\bibnamefont {Arima}}, \bibinfo
  {author} {\bibfnamefont {D.}~\bibnamefont {Okuyama}}, \bibinfo {author}
  {\bibfnamefont {K.}~\bibnamefont {Ohoyama}}, \bibinfo {author} {\bibfnamefont
  {S.}~\bibnamefont {Wakimoto}}, \bibinfo {author} {\bibfnamefont
  {K.}~\bibnamefont {Kakurai}}, \bibinfo {author} {\bibfnamefont
  {S.}~\bibnamefont {Ishiwata}},\ and\ \bibinfo {author} {\bibfnamefont
  {Y.}~\bibnamefont {Tokura}},\ }\bibfield  {title} {\bibinfo {title} {Large
  topological {Hall} effect in a short-period helimagnet {MnGe}.},\ }\href
  {https://doi.org/10.1103/PhysRevLett.106.156603} {\bibfield  {journal}
  {\bibinfo  {journal} {Phys. Rev. Lett}\ }\textbf {\bibinfo {volume} {106}},\
  \bibinfo {pages} {156603} (\bibinfo {year} {2011})}\BibitemShut {NoStop}%
\bibitem [{\citenamefont {Nakazawa}\ \emph {et~al.}(2018)\citenamefont
  {Nakazawa}, \citenamefont {Bibes},\ and\ \citenamefont
  {Kohno}}]{Nakazawa:2018}%
  \BibitemOpen
  \bibfield  {author} {\bibinfo {author} {\bibfnamefont {K.}~\bibnamefont
  {Nakazawa}}, \bibinfo {author} {\bibfnamefont {M.}~\bibnamefont {Bibes}},\
  and\ \bibinfo {author} {\bibfnamefont {H.}~\bibnamefont {Kohno}},\ }\bibfield
   {title} {\bibinfo {title} {Topological {Hall} effect from strong to weak
  coupling.},\ }\href {https://doi.org/10.7566/JPSJ.87.033705} {\bibfield
  {journal} {\bibinfo  {journal} {J. Phys. Soc. Jpn}\ }\textbf {\bibinfo
  {volume} {87}},\ \bibinfo {pages} {033705} (\bibinfo {year}
  {2018})}\BibitemShut {NoStop}%
\bibitem [{\citenamefont {Taguchi}\ \emph {et~al.}(2001)\citenamefont
  {Taguchi}, \citenamefont {Oohara}, \citenamefont {Yoshizawa}, \citenamefont
  {Nagaosa},\ and\ \citenamefont {Tokura}}]{Taguchi:2001}%
  \BibitemOpen
  \bibfield  {author} {\bibinfo {author} {\bibfnamefont {Y.}~\bibnamefont
  {Taguchi}}, \bibinfo {author} {\bibfnamefont {Y.}~\bibnamefont {Oohara}},
  \bibinfo {author} {\bibfnamefont {H.}~\bibnamefont {Yoshizawa}}, \bibinfo
  {author} {\bibfnamefont {N.}~\bibnamefont {Nagaosa}},\ and\ \bibinfo {author}
  {\bibfnamefont {Y.}~\bibnamefont {Tokura}},\ }\bibfield  {title} {\bibinfo
  {title} {Spin chirality, {Berry} phase, and anomalous {Hall} effect in a
  frustrated ferromagnet.},\ }\href {https://doi.org/10.1126/science.1058161}
  {\bibfield  {journal} {\bibinfo  {journal} {Science}\ }\textbf {\bibinfo
  {volume} {291}},\ \bibinfo {pages} {2573} (\bibinfo {year}
  {2001})}\BibitemShut {NoStop}%
\bibitem [{\citenamefont {Spencer}\ \emph {et~al.}(2018)\citenamefont
  {Spencer}, \citenamefont {Gayles}, \citenamefont {Porter}, \citenamefont
  {Sugimoto}, \citenamefont {Aslam}, \citenamefont {Kinane}, \citenamefont
  {Charlton}, \citenamefont {Freimuth}, \citenamefont {Chadov}, \citenamefont
  {Langridge}, \citenamefont {Sinova}, \citenamefont {Felser}, \citenamefont
  {Bl\"{u}gel}, \citenamefont {Mokrousov},\ and\ \citenamefont
  {Marrows}}]{Spencer:2018}%
  \BibitemOpen
  \bibfield  {author} {\bibinfo {author} {\bibfnamefont {C.~S.}\ \bibnamefont
  {Spencer}}, \bibinfo {author} {\bibfnamefont {J.}~\bibnamefont {Gayles}},
  \bibinfo {author} {\bibfnamefont {N.~A.}\ \bibnamefont {Porter}}, \bibinfo
  {author} {\bibfnamefont {S.}~\bibnamefont {Sugimoto}}, \bibinfo {author}
  {\bibfnamefont {Z.}~\bibnamefont {Aslam}}, \bibinfo {author} {\bibfnamefont
  {C.~J.}\ \bibnamefont {Kinane}}, \bibinfo {author} {\bibfnamefont {T.~R.}\
  \bibnamefont {Charlton}}, \bibinfo {author} {\bibfnamefont {F.}~\bibnamefont
  {Freimuth}}, \bibinfo {author} {\bibfnamefont {S.}~\bibnamefont {Chadov}},
  \bibinfo {author} {\bibfnamefont {S.}~\bibnamefont {Langridge}}, \bibinfo
  {author} {\bibfnamefont {J.}~\bibnamefont {Sinova}}, \bibinfo {author}
  {\bibfnamefont {C.}~\bibnamefont {Felser}}, \bibinfo {author} {\bibfnamefont
  {S.}~\bibnamefont {Bl\"{u}gel}}, \bibinfo {author} {\bibfnamefont
  {Y.}~\bibnamefont {Mokrousov}},\ and\ \bibinfo {author} {\bibfnamefont
  {C.~H.}\ \bibnamefont {Marrows}},\ }\bibfield  {title} {\bibinfo {title}
  {Helical magnetic structure and the anomalous and topological {Hall} effects
  in epitaxial {B20} {Fe$_{(1-\mathrm{y})}$Co$_{\mathrm{y}}$Ge} films},\ }\href
  {https://doi.org/10.1103/PhysRevB.97.214406} {\bibfield  {journal} {\bibinfo
  {journal} {Phys. Rev. B}\ }\textbf {\bibinfo {volume} {97}},\ \bibinfo
  {pages} {214406} (\bibinfo {year} {2018})}\BibitemShut {NoStop}%
\bibitem [{\citenamefont {Neubauer}\ \emph {et~al.}(2009)\citenamefont
  {Neubauer}, \citenamefont {Pfleiderer}, \citenamefont {Binz}, \citenamefont
  {Rosch}, \citenamefont {Ritz}, \citenamefont {Niklowitz},\ and\ \citenamefont
  {B\"oni}}]{Neubauer:2009}%
  \BibitemOpen
  \bibfield  {author} {\bibinfo {author} {\bibfnamefont {A.}~\bibnamefont
  {Neubauer}}, \bibinfo {author} {\bibfnamefont {C.}~\bibnamefont
  {Pfleiderer}}, \bibinfo {author} {\bibfnamefont {B.}~\bibnamefont {Binz}},
  \bibinfo {author} {\bibfnamefont {A.}~\bibnamefont {Rosch}}, \bibinfo
  {author} {\bibfnamefont {R.}~\bibnamefont {Ritz}}, \bibinfo {author}
  {\bibfnamefont {P.~G.}\ \bibnamefont {Niklowitz}},\ and\ \bibinfo {author}
  {\bibfnamefont {P.}~\bibnamefont {B\"oni}},\ }\bibfield  {title} {\bibinfo
  {title} {Topological {Hall} effect in the {$A$} phase of {MnSi}.},\ }\href
  {https://doi.org/10.1103/PhysRevLett.102.186602} {\bibfield  {journal}
  {\bibinfo  {journal} {Phys. Rev. Lett.}\ }\textbf {\bibinfo {volume} {102}},\
  \bibinfo {pages} {186602} (\bibinfo {year} {2009})}\BibitemShut {NoStop}%
\bibitem [{\citenamefont {Porter}\ \emph {et~al.}(2014)\citenamefont {Porter},
  \citenamefont {Gartside},\ and\ \citenamefont {Marrows}}]{Porter:2014}%
  \BibitemOpen
  \bibfield  {author} {\bibinfo {author} {\bibfnamefont {N.~A.}\ \bibnamefont
  {Porter}}, \bibinfo {author} {\bibfnamefont {J.~C.}\ \bibnamefont
  {Gartside}},\ and\ \bibinfo {author} {\bibfnamefont {C.~H.}\ \bibnamefont
  {Marrows}},\ }\bibfield  {title} {\bibinfo {title} {Scattering mechanisms in
  textured {FeGe} thin films: Magnetoresistance and the anomalous {Hall}
  effect.},\ }\href {https://doi.org/10.1103/PhysRevB.90.024403} {\bibfield
  {journal} {\bibinfo  {journal} {Phys. Rev. B}\ }\textbf {\bibinfo {volume}
  {90}},\ \bibinfo {pages} {024403} (\bibinfo {year} {2014})}\BibitemShut
  {NoStop}%
\bibitem [{\citenamefont {Kurumaji}\ \emph {et~al.}(2019)\citenamefont
  {Kurumaji}, \citenamefont {Nakajima}, \citenamefont {Hirschberger},
  \citenamefont {Kikkawa}, \citenamefont {Yamasaki}, \citenamefont {Sagayama},
  \citenamefont {Nakao}, \citenamefont {Taguchi}, \citenamefont {Arima},\ and\
  \citenamefont {Tokura}}]{Kurumaji:2019}%
  \BibitemOpen
  \bibfield  {author} {\bibinfo {author} {\bibfnamefont {T.}~\bibnamefont
  {Kurumaji}}, \bibinfo {author} {\bibfnamefont {T.}~\bibnamefont {Nakajima}},
  \bibinfo {author} {\bibfnamefont {M.}~\bibnamefont {Hirschberger}}, \bibinfo
  {author} {\bibfnamefont {A.}~\bibnamefont {Kikkawa}}, \bibinfo {author}
  {\bibfnamefont {Y.}~\bibnamefont {Yamasaki}}, \bibinfo {author}
  {\bibfnamefont {H.}~\bibnamefont {Sagayama}}, \bibinfo {author}
  {\bibfnamefont {H.}~\bibnamefont {Nakao}}, \bibinfo {author} {\bibfnamefont
  {Y.}~\bibnamefont {Taguchi}}, \bibinfo {author} {\bibfnamefont
  {T.}~\bibnamefont {Arima}},\ and\ \bibinfo {author} {\bibfnamefont
  {Y.}~\bibnamefont {Tokura}},\ }\bibfield  {title} {\bibinfo {title} {Skyrmion
  lattice with a giant topological {Hall} effect in a frustrated
  triangular-lattice magnet.},\ }\href
  {https://science.sciencemag.org/content/365/6456/914} {\bibfield  {journal}
  {\bibinfo  {journal} {Science}\ }\textbf {\bibinfo {volume} {365}},\ \bibinfo
  {pages} {914} (\bibinfo {year} {2019})}\BibitemShut {NoStop}%
\bibitem [{\citenamefont {Schulz}\ \emph {et~al.}(2012)\citenamefont {Schulz},
  \citenamefont {Ritz}, \citenamefont {Bauer}, \citenamefont {Halder},
  \citenamefont {Wagner}, \citenamefont {Franz}, \citenamefont {Pfleiderer},
  \citenamefont {Everschor}, \citenamefont {Garst},\ and\ \citenamefont
  {Rosch}}]{Schulz:2012}%
  \BibitemOpen
  \bibfield  {author} {\bibinfo {author} {\bibfnamefont {T.}~\bibnamefont
  {Schulz}}, \bibinfo {author} {\bibfnamefont {R.}~\bibnamefont {Ritz}},
  \bibinfo {author} {\bibfnamefont {A.}~\bibnamefont {Bauer}}, \bibinfo
  {author} {\bibfnamefont {M.}~\bibnamefont {Halder}}, \bibinfo {author}
  {\bibfnamefont {M.}~\bibnamefont {Wagner}}, \bibinfo {author} {\bibfnamefont
  {C.}~\bibnamefont {Franz}}, \bibinfo {author} {\bibfnamefont
  {C.}~\bibnamefont {Pfleiderer}}, \bibinfo {author} {\bibfnamefont
  {K.}~\bibnamefont {Everschor}}, \bibinfo {author} {\bibfnamefont
  {M.}~\bibnamefont {Garst}},\ and\ \bibinfo {author} {\bibfnamefont
  {A.}~\bibnamefont {Rosch}},\ }\bibfield  {title} {\bibinfo {title} {Emergent
  electrodynamics of skyrmions in a chiral magnet.},\ }\href
  {https://doi.org/10.1038/nphys2231} {\bibfield  {journal} {\bibinfo
  {journal} {Nat. Phys.}\ }\textbf {\bibinfo {volume} {8}},\ \bibinfo {pages}
  {301} (\bibinfo {year} {2012})}\BibitemShut {NoStop}%
\bibitem [{\citenamefont {Wang}\ \emph {et~al.}(2020)\citenamefont {Wang},
  \citenamefont {Zeng}, \citenamefont {Liu}, \citenamefont {Zhang},
  \citenamefont {Ma}, \citenamefont {Xu}, \citenamefont {Liang}, \citenamefont
  {Zhang}, \citenamefont {Wu}, \citenamefont {Che}, \citenamefont {Han},\ and\
  \citenamefont {Huang}}]{Wang:2020}%
  \BibitemOpen
  \bibfield  {author} {\bibinfo {author} {\bibfnamefont {S.}~\bibnamefont
  {Wang}}, \bibinfo {author} {\bibfnamefont {Q.}~\bibnamefont {Zeng}}, \bibinfo
  {author} {\bibfnamefont {D.}~\bibnamefont {Liu}}, \bibinfo {author}
  {\bibfnamefont {H.}~\bibnamefont {Zhang}}, \bibinfo {author} {\bibfnamefont
  {L.}~\bibnamefont {Ma}}, \bibinfo {author} {\bibfnamefont {G.}~\bibnamefont
  {Xu}}, \bibinfo {author} {\bibfnamefont {Y.}~\bibnamefont {Liang}}, \bibinfo
  {author} {\bibfnamefont {Z.}~\bibnamefont {Zhang}}, \bibinfo {author}
  {\bibfnamefont {H.}~\bibnamefont {Wu}}, \bibinfo {author} {\bibfnamefont
  {R.}~\bibnamefont {Che}}, \bibinfo {author} {\bibfnamefont {X.}~\bibnamefont
  {Han}},\ and\ \bibinfo {author} {\bibfnamefont {Q.}~\bibnamefont {Huang}},\
  }\bibfield  {title} {\bibinfo {title} {Giant topological {Hall} effect and
  superstable spontaneous skyrmions below 330 {K} in a centrosymmetric complex
  noncollinear ferromagnet {NdMn$_2$Ge$_2$}.},\ }\href
  {https://pubs.acs.org/doi/10.1021/acsami.0c04632} {\bibfield  {journal}
  {\bibinfo  {journal} {ACS Appl. Mater. Interfaces}\ }\textbf {\bibinfo
  {volume} {12}},\ \bibinfo {pages} {24125} (\bibinfo {year}
  {2020})}\BibitemShut {NoStop}%
\bibitem [{\citenamefont {Vistoli}\ \emph {et~al.}(2019)\citenamefont
  {Vistoli}, \citenamefont {Wang}, \citenamefont {Sander}, \citenamefont {Zhu},
  \citenamefont {Casals}, \citenamefont {Cichelero}, \citenamefont
  {Barth{\'e}l{\'e}my}, \citenamefont {Fusil}, \citenamefont {Herranz},
  \citenamefont {Valencia} \emph {et~al.}}]{Vistoli:2019giant}%
  \BibitemOpen
  \bibfield  {author} {\bibinfo {author} {\bibfnamefont {L.}~\bibnamefont
  {Vistoli}}, \bibinfo {author} {\bibfnamefont {W.}~\bibnamefont {Wang}},
  \bibinfo {author} {\bibfnamefont {A.}~\bibnamefont {Sander}}, \bibinfo
  {author} {\bibfnamefont {Q.}~\bibnamefont {Zhu}}, \bibinfo {author}
  {\bibfnamefont {B.}~\bibnamefont {Casals}}, \bibinfo {author} {\bibfnamefont
  {R.}~\bibnamefont {Cichelero}}, \bibinfo {author} {\bibfnamefont
  {A.}~\bibnamefont {Barth{\'e}l{\'e}my}}, \bibinfo {author} {\bibfnamefont
  {S.}~\bibnamefont {Fusil}}, \bibinfo {author} {\bibfnamefont
  {G.}~\bibnamefont {Herranz}}, \bibinfo {author} {\bibfnamefont
  {S.}~\bibnamefont {Valencia}}, \emph {et~al.},\ }\bibfield  {title} {\bibinfo
  {title} {Giant topological {Hall} effect in correlated oxide thin films.},\
  }\href {https://doi.org/10.1038/s41567-018-0307-5} {\bibfield  {journal}
  {\bibinfo  {journal} {Nat. Phys.}\ }\textbf {\bibinfo {volume} {15}},\
  \bibinfo {pages} {67} (\bibinfo {year} {2019})}\BibitemShut {NoStop}%
\bibitem [{\citenamefont {Togawa}\ \emph {et~al.}(2012)\citenamefont {Togawa},
  \citenamefont {Koyama}, \citenamefont {Takayanagi}, \citenamefont {Mori},
  \citenamefont {Kousaka}, \citenamefont {Akimitsu}, \citenamefont {Nishihara},
  \citenamefont {Inoue}, \citenamefont {Ovchinnikov},\ and\ \citenamefont
  {Kishine}}]{Togawa:2012}%
  \BibitemOpen
  \bibfield  {author} {\bibinfo {author} {\bibfnamefont {Y.}~\bibnamefont
  {Togawa}}, \bibinfo {author} {\bibfnamefont {T.}~\bibnamefont {Koyama}},
  \bibinfo {author} {\bibfnamefont {K.}~\bibnamefont {Takayanagi}}, \bibinfo
  {author} {\bibfnamefont {S.}~\bibnamefont {Mori}}, \bibinfo {author}
  {\bibfnamefont {Y.}~\bibnamefont {Kousaka}}, \bibinfo {author} {\bibfnamefont
  {J.}~\bibnamefont {Akimitsu}}, \bibinfo {author} {\bibfnamefont
  {S.}~\bibnamefont {Nishihara}}, \bibinfo {author} {\bibfnamefont
  {K.}~\bibnamefont {Inoue}}, \bibinfo {author} {\bibfnamefont {A.~S.}\
  \bibnamefont {Ovchinnikov}},\ and\ \bibinfo {author} {\bibfnamefont
  {J.}~\bibnamefont {Kishine}},\ }\bibfield  {title} {\bibinfo {title} {Chiral
  magnetic soliton lattice on a chiral helimagnet.},\ }\href
  {https://doi.org/10.1103/PhysRevLett.108.107202} {\bibfield  {journal}
  {\bibinfo  {journal} {Phys. Rev. Lett.}\ }\textbf {\bibinfo {volume} {108}},\
  \bibinfo {pages} {107202} (\bibinfo {year} {2012})}\BibitemShut {NoStop}%
\bibitem [{\citenamefont {Parkin}\ and\ \citenamefont
  {Friend}(1980{\natexlab{a}})}]{Parkin:1980a}%
  \BibitemOpen
  \bibfield  {author} {\bibinfo {author} {\bibfnamefont {S.~S.~P.}\
  \bibnamefont {Parkin}}\ and\ \bibinfo {author} {\bibfnamefont {R.~H.}\
  \bibnamefont {Friend}},\ }\bibfield  {title} {\bibinfo {title} {3d
  transition-metal intercalates of the niobium and tantalum dichalcogenides.
  {I}. magnetic properties.},\ }\href
  {https://doi.org/10.1080/13642818008245370} {\bibfield  {journal} {\bibinfo
  {journal} {Philos. Mag. B}\ }\textbf {\bibinfo {volume} {41}},\ \bibinfo
  {pages} {65} (\bibinfo {year} {1980}{\natexlab{a}})}\BibitemShut {NoStop}%
\bibitem [{\citenamefont {Parkin}\ and\ \citenamefont
  {Friend}(1980{\natexlab{b}})}]{Parkin:1980b}%
  \BibitemOpen
  \bibfield  {author} {\bibinfo {author} {\bibfnamefont {S.~S.~P.}\
  \bibnamefont {Parkin}}\ and\ \bibinfo {author} {\bibfnamefont {R.~H.}\
  \bibnamefont {Friend}},\ }\bibfield  {title} {\bibinfo {title} {3d
  transition-metal intercalates of the niobium and tantalum dichalcogenides.
  {II}. transport properties.},\ }\href
  {https://doi.org/10.1080/13642818008245371} {\bibfield  {journal} {\bibinfo
  {journal} {Philos. Mag. B}\ }\textbf {\bibinfo {volume} {41}},\ \bibinfo
  {pages} {95} (\bibinfo {year} {1980}{\natexlab{b}})}\BibitemShut {NoStop}%
\bibitem [{\citenamefont {Cao}\ \emph {et~al.}(2020)\citenamefont {Cao},
  \citenamefont {Huang}, \citenamefont {Yin}, \citenamefont {Xie},
  \citenamefont {Liu}, \citenamefont {Wang}, \citenamefont {Zhu}, \citenamefont
  {Mandrus}, \citenamefont {Wang},\ and\ \citenamefont {Huang}}]{Cao:2020}%
  \BibitemOpen
  \bibfield  {author} {\bibinfo {author} {\bibfnamefont {Y.}~\bibnamefont
  {Cao}}, \bibinfo {author} {\bibfnamefont {Z.}~\bibnamefont {Huang}}, \bibinfo
  {author} {\bibfnamefont {Y.}~\bibnamefont {Yin}}, \bibinfo {author}
  {\bibfnamefont {H.}~\bibnamefont {Xie}}, \bibinfo {author} {\bibfnamefont
  {B.}~\bibnamefont {Liu}}, \bibinfo {author} {\bibfnamefont {B.}~\bibnamefont
  {Wang}}, \bibinfo {author} {\bibfnamefont {C.}~\bibnamefont {Zhu}}, \bibinfo
  {author} {\bibfnamefont {D.}~\bibnamefont {Mandrus}}, \bibinfo {author}
  {\bibfnamefont {L.}~\bibnamefont {Wang}},\ and\ \bibinfo {author}
  {\bibfnamefont {W.}~\bibnamefont {Huang}},\ }\bibfield  {title} {\bibinfo
  {title} {Overview and advance in a layered chircal helimagnet
  {Cr$_{1/3}$NbS$_{2}$}.},\ }\href
  {https://doi.org/10.1016/j.mtadv.2020.100080} {\bibfield  {journal} {\bibinfo
   {journal} {Mater. Today Adv.}\ }\textbf {\bibinfo {volume} {7}},\ \bibinfo
  {pages} {100080} (\bibinfo {year} {2020})}\BibitemShut {NoStop}%
\bibitem [{\citenamefont {Togawa}\ \emph {et~al.}(2015)\citenamefont {Togawa},
  \citenamefont {Koyama}, \citenamefont {Nishimori}, \citenamefont {Matsumoto},
  \citenamefont {McVitie}, \citenamefont {McGrouther}, \citenamefont {Stamps},
  \citenamefont {Kousaka}, \citenamefont {Akimitsu}, \citenamefont {Nishihara},
  \citenamefont {Inoue}, \citenamefont {Bostrem}, \citenamefont {Sinitsyn},
  \citenamefont {Ovchinnikov},\ and\ \citenamefont {Kishine}}]{Togawa:2015}%
  \BibitemOpen
  \bibfield  {author} {\bibinfo {author} {\bibfnamefont {Y.}~\bibnamefont
  {Togawa}}, \bibinfo {author} {\bibfnamefont {T.}~\bibnamefont {Koyama}},
  \bibinfo {author} {\bibfnamefont {Y.}~\bibnamefont {Nishimori}}, \bibinfo
  {author} {\bibfnamefont {Y.}~\bibnamefont {Matsumoto}}, \bibinfo {author}
  {\bibfnamefont {S.}~\bibnamefont {McVitie}}, \bibinfo {author} {\bibfnamefont
  {D.}~\bibnamefont {McGrouther}}, \bibinfo {author} {\bibfnamefont {R.~L.}\
  \bibnamefont {Stamps}}, \bibinfo {author} {\bibfnamefont {Y.}~\bibnamefont
  {Kousaka}}, \bibinfo {author} {\bibfnamefont {J.}~\bibnamefont {Akimitsu}},
  \bibinfo {author} {\bibfnamefont {S.}~\bibnamefont {Nishihara}}, \bibinfo
  {author} {\bibfnamefont {K.}~\bibnamefont {Inoue}}, \bibinfo {author}
  {\bibfnamefont {I.~G.}\ \bibnamefont {Bostrem}}, \bibinfo {author}
  {\bibfnamefont {V.~E.}\ \bibnamefont {Sinitsyn}}, \bibinfo {author}
  {\bibfnamefont {A.~S.}\ \bibnamefont {Ovchinnikov}},\ and\ \bibinfo {author}
  {\bibfnamefont {J.}~\bibnamefont {Kishine}},\ }\bibfield  {title} {\bibinfo
  {title} {Magnetic soliton confinement and discretization effects arising from
  macroscopic coherence in a chiral spin soliton lattice.},\ }\href
  {https://doi.org/10.1103/PhysRevB.92.220412} {\bibfield  {journal} {\bibinfo
  {journal} {Phys. Rev. B}\ }\textbf {\bibinfo {volume} {92}},\ \bibinfo
  {pages} {220412(R)} (\bibinfo {year} {2015})}\BibitemShut {NoStop}%
\bibitem [{\citenamefont {Yonemura}\ \emph {et~al.}(2017)\citenamefont
  {Yonemura}, \citenamefont {Shimamoto}, \citenamefont {Kida}, \citenamefont
  {Yoshizawa}, \citenamefont {Kousaka}, \citenamefont {Nishihara},
  \citenamefont {Goncalves}, \citenamefont {Akimitsu}, \citenamefont {Inoue},
  \citenamefont {Hagiwara},\ and\ \citenamefont {Togawa}}]{Yonemure:2017}%
  \BibitemOpen
  \bibfield  {author} {\bibinfo {author} {\bibfnamefont {J.-I.}\ \bibnamefont
  {Yonemura}}, \bibinfo {author} {\bibfnamefont {Y.}~\bibnamefont {Shimamoto}},
  \bibinfo {author} {\bibfnamefont {T.}~\bibnamefont {Kida}}, \bibinfo {author}
  {\bibfnamefont {D.}~\bibnamefont {Yoshizawa}}, \bibinfo {author}
  {\bibfnamefont {Y.}~\bibnamefont {Kousaka}}, \bibinfo {author} {\bibfnamefont
  {S.}~\bibnamefont {Nishihara}}, \bibinfo {author} {\bibfnamefont {F.~J.~T.}\
  \bibnamefont {Goncalves}}, \bibinfo {author} {\bibfnamefont {J.}~\bibnamefont
  {Akimitsu}}, \bibinfo {author} {\bibfnamefont {K.}~\bibnamefont {Inoue}},
  \bibinfo {author} {\bibfnamefont {M.}~\bibnamefont {Hagiwara}},\ and\
  \bibinfo {author} {\bibfnamefont {Y.}~\bibnamefont {Togawa}},\ }\bibfield
  {title} {\bibinfo {title} {Magnetic solitons and magnetic phase diagram of
  the hexagonal chiral crystal {CrNb$_{3}$S$_{6}$} in oblique magnetic
  fields.},\ }\href {https://doi.org/10.1103/PhysRevB.96.184423} {\bibfield
  {journal} {\bibinfo  {journal} {Phys. Rev. B}\ }\textbf {\bibinfo {volume}
  {96}},\ \bibinfo {pages} {184423} (\bibinfo {year} {2017})}\BibitemShut
  {NoStop}%
\bibitem [{SMn()}]{SMnote}%
  \BibitemOpen
  \href@noop {} {}\bibinfo {note} {See Supplemental Material at [URL will be
  inserted by publisher] for methodological details and supporting experimental
  results and theoretical simulations.}\BibitemShut {Stop}%
\bibitem [{\citenamefont {Colvin}\ \emph {et~al.}(1960)\citenamefont {Colvin},
  \citenamefont {Legvold},\ and\ \citenamefont {Spedding}}]{Colvin:1960}%
  \BibitemOpen
  \bibfield  {author} {\bibinfo {author} {\bibfnamefont {R.~V.}\ \bibnamefont
  {Colvin}}, \bibinfo {author} {\bibfnamefont {S.}~\bibnamefont {Legvold}},\
  and\ \bibinfo {author} {\bibfnamefont {F.~H.}\ \bibnamefont {Spedding}},\
  }\bibfield  {title} {\bibinfo {title} {Electrical resistivity of the heavy
  rare-earth metals.},\ }\href {https://doi.org/10.1103/PhysRev.120.741}
  {\bibfield  {journal} {\bibinfo  {journal} {Phys. Rev.}\ }\textbf {\bibinfo
  {volume} {120}},\ \bibinfo {pages} {741} (\bibinfo {year}
  {1960})}\BibitemShut {NoStop}%
\bibitem [{\citenamefont {Barati}\ \emph {et~al.}(1993)\citenamefont {Barati},
  \citenamefont {Datars}, \citenamefont {Chien}, \citenamefont {Stager},\ and\
  \citenamefont {Garrett}}]{Barati:1993}%
  \BibitemOpen
  \bibfield  {author} {\bibinfo {author} {\bibfnamefont {M.}~\bibnamefont
  {Barati}}, \bibinfo {author} {\bibfnamefont {W.~R.}\ \bibnamefont {Datars}},
  \bibinfo {author} {\bibfnamefont {T.~R.}\ \bibnamefont {Chien}}, \bibinfo
  {author} {\bibfnamefont {C.~V.}\ \bibnamefont {Stager}},\ and\ \bibinfo
  {author} {\bibfnamefont {J.~D.}\ \bibnamefont {Garrett}},\ }\bibfield
  {title} {\bibinfo {title} {Resistivity, {Hall} effect, and magnetic
  susceptibility of {${\mathrm{UPd}}_{2}{\mathrm{Si}}_{2}$}.},\ }\href
  {https://doi.org/10.1103/PhysRevB.48.16926} {\bibfield  {journal} {\bibinfo
  {journal} {Phys. Rev. B}\ }\textbf {\bibinfo {volume} {48}},\ \bibinfo
  {pages} {16926} (\bibinfo {year} {1993})}\BibitemShut {NoStop}%
\bibitem [{\citenamefont {Haas}(1968)}]{Haas:1968}%
  \BibitemOpen
  \bibfield  {author} {\bibinfo {author} {\bibfnamefont {C.}~\bibnamefont
  {Haas}},\ }\bibfield  {title} {\bibinfo {title} {Spin-disorder scattering and
  magnetoresistance of magnetic semiconductors.},\ }\href
  {https://doi.org/10.1103/PhysRev.168.531} {\bibfield  {journal} {\bibinfo
  {journal} {Phys. Rev.}\ }\textbf {\bibinfo {volume} {168}},\ \bibinfo {pages}
  {531} (\bibinfo {year} {1968})}\BibitemShut {NoStop}%
\bibitem [{\citenamefont {Ghimire}\ \emph {et~al.}(2013)\citenamefont
  {Ghimire}, \citenamefont {McGuire}, \citenamefont {Parker}, \citenamefont
  {Sipos}, \citenamefont {Tang}, \citenamefont {Yan}, \citenamefont {Sales},\
  and\ \citenamefont {Mandrus}}]{Ghimire:2012}%
  \BibitemOpen
  \bibfield  {author} {\bibinfo {author} {\bibfnamefont {N.~J.}\ \bibnamefont
  {Ghimire}}, \bibinfo {author} {\bibfnamefont {M.~A.}\ \bibnamefont
  {McGuire}}, \bibinfo {author} {\bibfnamefont {D.~S.}\ \bibnamefont {Parker}},
  \bibinfo {author} {\bibfnamefont {B.}~\bibnamefont {Sipos}}, \bibinfo
  {author} {\bibfnamefont {S.}~\bibnamefont {Tang}}, \bibinfo {author}
  {\bibfnamefont {J.-Q.}\ \bibnamefont {Yan}}, \bibinfo {author} {\bibfnamefont
  {B.~C.}\ \bibnamefont {Sales}},\ and\ \bibinfo {author} {\bibfnamefont
  {D.}~\bibnamefont {Mandrus}},\ }\bibfield  {title} {\bibinfo {title}
  {Magnetic phase transition in single crystals of the chiral helimagnet
  {Cr$_{1/3}$NbS$_{2}$}.},\ }\href
  {https://link.aps.org/doi/10.1103/PhysRevB.87.104403} {\bibfield  {journal}
  {\bibinfo  {journal} {Phys. Rev. B}\ }\textbf {\bibinfo {volume} {87}},\
  \bibinfo {pages} {104403} (\bibinfo {year} {2013})}\BibitemShut {NoStop}%
\bibitem [{\citenamefont {Dyadkin}\ \emph {et~al.}(2015)\citenamefont
  {Dyadkin}, \citenamefont {Mushenok}, \citenamefont {Bosak}, \citenamefont
  {Menzel}, \citenamefont {Grigoriev}, \citenamefont {Pattison},\ and\
  \citenamefont {Chernyshov}}]{Dyadkin:2015}%
  \BibitemOpen
  \bibfield  {author} {\bibinfo {author} {\bibfnamefont {V.}~\bibnamefont
  {Dyadkin}}, \bibinfo {author} {\bibfnamefont {F.}~\bibnamefont {Mushenok}},
  \bibinfo {author} {\bibfnamefont {A.}~\bibnamefont {Bosak}}, \bibinfo
  {author} {\bibfnamefont {D.}~\bibnamefont {Menzel}}, \bibinfo {author}
  {\bibfnamefont {S.}~\bibnamefont {Grigoriev}}, \bibinfo {author}
  {\bibfnamefont {P.}~\bibnamefont {Pattison}},\ and\ \bibinfo {author}
  {\bibfnamefont {D.}~\bibnamefont {Chernyshov}},\ }\bibfield  {title}
  {\bibinfo {title} {Structural disorder versus chiral magnetism in
  {Cr$_{1/3}$NbS$_{2}$}.},\ }\href {https://doi.org/10.1103/PhysRevB.91.184205}
  {\bibfield  {journal} {\bibinfo  {journal} {Phys. Rev. B}\ }\textbf {\bibinfo
  {volume} {91}},\ \bibinfo {pages} {184205} (\bibinfo {year}
  {2015})}\BibitemShut {NoStop}%
\bibitem [{\citenamefont {Lee}\ \emph {et~al.}(2007)\citenamefont {Lee},
  \citenamefont {Onose}, \citenamefont {Tokura},\ and\ \citenamefont
  {Ong}}]{Lee:2007}%
  \BibitemOpen
  \bibfield  {author} {\bibinfo {author} {\bibfnamefont {M.}~\bibnamefont
  {Lee}}, \bibinfo {author} {\bibfnamefont {Y.}~\bibnamefont {Onose}}, \bibinfo
  {author} {\bibfnamefont {Y.}~\bibnamefont {Tokura}},\ and\ \bibinfo {author}
  {\bibfnamefont {N.~P.}\ \bibnamefont {Ong}},\ }\bibfield  {title} {\bibinfo
  {title} {Hidden constant in the anomalous {Hall} effect of high-purity magnet
  {MnSi}.},\ }\href {https://doi.org/10.1103/PhysRevB.75.172403} {\bibfield
  {journal} {\bibinfo  {journal} {Phys. Rev. B}\ }\textbf {\bibinfo {volume}
  {75}},\ \bibinfo {pages} {172403} (\bibinfo {year} {2007})}\BibitemShut
  {NoStop}%
\bibitem [{\citenamefont {Chapman}\ \emph {et~al.}(2013)\citenamefont
  {Chapman}, \citenamefont {Grossnickle}, \citenamefont {Wolf},\ and\
  \citenamefont {Lee}}]{Chapman:2013}%
  \BibitemOpen
  \bibfield  {author} {\bibinfo {author} {\bibfnamefont {B.~J.}\ \bibnamefont
  {Chapman}}, \bibinfo {author} {\bibfnamefont {M.~G.}\ \bibnamefont
  {Grossnickle}}, \bibinfo {author} {\bibfnamefont {T.}~\bibnamefont {Wolf}},\
  and\ \bibinfo {author} {\bibfnamefont {M.}~\bibnamefont {Lee}},\ }\bibfield
  {title} {\bibinfo {title} {Large enhancement of emergent magnetic fields in
  {MnSi} with impurities and pressure},\ }\href
  {https://doi.org/10.1103/PhysRevB.88.214406} {\bibfield  {journal} {\bibinfo
  {journal} {Phys. Rev. B}\ }\textbf {\bibinfo {volume} {88}},\ \bibinfo
  {pages} {214406} (\bibinfo {year} {2013})}\BibitemShut {NoStop}%
\bibitem [{\citenamefont {Zhang}\ \emph {et~al.}(2020)\citenamefont {Zhang},
  \citenamefont {Huang}, \citenamefont {Hao}, \citenamefont {Yang},
  \citenamefont {Noordhoek}, \citenamefont {Pandey}, \citenamefont {Zhou},\
  and\ \citenamefont {Liu}}]{Zhang:2020}%
  \BibitemOpen
  \bibfield  {author} {\bibinfo {author} {\bibfnamefont {H.}~\bibnamefont
  {Zhang}}, \bibinfo {author} {\bibfnamefont {Q.}~\bibnamefont {Huang}},
  \bibinfo {author} {\bibfnamefont {L.}~\bibnamefont {Hao}}, \bibinfo {author}
  {\bibfnamefont {J.}~\bibnamefont {Yang}}, \bibinfo {author} {\bibfnamefont
  {K.}~\bibnamefont {Noordhoek}}, \bibinfo {author} {\bibfnamefont
  {S.}~\bibnamefont {Pandey}}, \bibinfo {author} {\bibfnamefont
  {H.}~\bibnamefont {Zhou}},\ and\ \bibinfo {author} {\bibfnamefont
  {J.}~\bibnamefont {Liu}},\ }\bibfield  {title} {\bibinfo {title} {Anomalous
  magnetoresistance in centrosymmetric skyrmion-lattice magnet
  {Gd$_{2}$PdSi$_{3}$}.},\ }\href {https://doi.org/10.1088/1367-2630/aba650}
  {\bibfield  {journal} {\bibinfo  {journal} {New J. Phys.}\ }\textbf {\bibinfo
  {volume} {22}},\ \bibinfo {pages} {083056} (\bibinfo {year}
  {2020})}\BibitemShut {NoStop}%
\bibitem [{\citenamefont {Chang}\ and\ \citenamefont {Niu}(1996)}]{Chang:1996}%
  \BibitemOpen
  \bibfield  {author} {\bibinfo {author} {\bibfnamefont {M.-C.}\ \bibnamefont
  {Chang}}\ and\ \bibinfo {author} {\bibfnamefont {Q.}~\bibnamefont {Niu}},\
  }\bibfield  {title} {\bibinfo {title} {{Berry} phase, hyperorbits, and the
  {Hofstadter} spectrum: Semiclassical dynamics in magnetic {Bloch} bands.},\
  }\href {https://doi.org/10.1103/PhysRevB.53.7010} {\bibfield  {journal}
  {\bibinfo  {journal} {Phys. Rev. B.}\ }\textbf {\bibinfo {volume} {53}},\
  \bibinfo {pages} {7010} (\bibinfo {year} {1996})}\BibitemShut {NoStop}%
\bibitem [{\citenamefont {Nagaosa}\ \emph {et~al.}(2010)\citenamefont
  {Nagaosa}, \citenamefont {Sinova}, \citenamefont {Onoda}, \citenamefont
  {MacDonald},\ and\ \citenamefont {Ong}}]{Nagaosa:2010review}%
  \BibitemOpen
  \bibfield  {author} {\bibinfo {author} {\bibfnamefont {N.}~\bibnamefont
  {Nagaosa}}, \bibinfo {author} {\bibfnamefont {J.}~\bibnamefont {Sinova}},
  \bibinfo {author} {\bibfnamefont {S.}~\bibnamefont {Onoda}}, \bibinfo
  {author} {\bibfnamefont {A.~H.}\ \bibnamefont {MacDonald}},\ and\ \bibinfo
  {author} {\bibfnamefont {N.~P.}\ \bibnamefont {Ong}},\ }\bibfield  {title}
  {\bibinfo {title} {Anomalous {Hall} effect.},\ }\href
  {https://doi.org/10.1103/RevModPhys.82.1539} {\bibfield  {journal} {\bibinfo
  {journal} {Rev. Mod. Phys.}\ }\textbf {\bibinfo {volume} {82}},\ \bibinfo
  {pages} {1539} (\bibinfo {year} {2010})}\BibitemShut {NoStop}%
\bibitem [{\citenamefont {Smit}(1955)}]{Smit:1955}%
  \BibitemOpen
  \bibfield  {author} {\bibinfo {author} {\bibfnamefont {J.}~\bibnamefont
  {Smit}},\ }\bibfield  {title} {\bibinfo {title} {The spontaneous {Hall}
  effect in ferromagnetics {I}.},\ }\href
  {https://doi.org/https://doi.org/10.1016/S0031-8914(55)92596-9} {\bibfield
  {journal} {\bibinfo  {journal} {Physica}\ }\textbf {\bibinfo {volume} {21}},\
  \bibinfo {pages} {877} (\bibinfo {year} {1955})}\BibitemShut {NoStop}%
\bibitem [{\citenamefont {Berger}(1970)}]{Berger:1970}%
  \BibitemOpen
  \bibfield  {author} {\bibinfo {author} {\bibfnamefont {L.}~\bibnamefont
  {Berger}},\ }\bibfield  {title} {\bibinfo {title} {Side-jump mechanism for
  the {Hall} effect of ferromagnets.},\ }\href
  {https://doi.org/10.1103/PhysRevB.2.4559} {\bibfield  {journal} {\bibinfo
  {journal} {Phys. Rev. B}\ }\textbf {\bibinfo {volume} {2}},\ \bibinfo {pages}
  {4559} (\bibinfo {year} {1970})}\BibitemShut {NoStop}%
\bibitem [{\citenamefont {Ernst}\ \emph {et~al.}(2019)\citenamefont {Ernst},
  \citenamefont {Sahoo}, \citenamefont {Sun}, \citenamefont {Nayak},
  \citenamefont {M\"uchler}, \citenamefont {Nayak}, \citenamefont {Kumar},
  \citenamefont {Gayles}, \citenamefont {Markou}, \citenamefont {Fecher},\ and\
  \citenamefont {Felser}}]{Ernst:2019}%
  \BibitemOpen
  \bibfield  {author} {\bibinfo {author} {\bibfnamefont {B.}~\bibnamefont
  {Ernst}}, \bibinfo {author} {\bibfnamefont {R.}~\bibnamefont {Sahoo}},
  \bibinfo {author} {\bibfnamefont {Y.}~\bibnamefont {Sun}}, \bibinfo {author}
  {\bibfnamefont {J.}~\bibnamefont {Nayak}}, \bibinfo {author} {\bibfnamefont
  {L.}~\bibnamefont {M\"uchler}}, \bibinfo {author} {\bibfnamefont {A.~K.}\
  \bibnamefont {Nayak}}, \bibinfo {author} {\bibfnamefont {N.}~\bibnamefont
  {Kumar}}, \bibinfo {author} {\bibfnamefont {J.}~\bibnamefont {Gayles}},
  \bibinfo {author} {\bibfnamefont {A.}~\bibnamefont {Markou}}, \bibinfo
  {author} {\bibfnamefont {G.~H.}\ \bibnamefont {Fecher}},\ and\ \bibinfo
  {author} {\bibfnamefont {C.}~\bibnamefont {Felser}},\ }\bibfield  {title}
  {\bibinfo {title} {Anomalous {Hall} effect and the role of {Berry} curvature
  in {Co$_{2}$TiSn} {Heusler} films.},\ }\href
  {https://doi.org/10.1103/PhysRevB.100.054445} {\bibfield  {journal} {\bibinfo
   {journal} {Phys. Rev. B}\ }\textbf {\bibinfo {volume} {100}},\ \bibinfo
  {pages} {054445} (\bibinfo {year} {2019})}\BibitemShut {NoStop}%
\bibitem [{\citenamefont {Bruno}\ \emph {et~al.}(2004)\citenamefont {Bruno},
  \citenamefont {Dugaev},\ and\ \citenamefont {Taillefumier}}]{Bruno:2004}%
  \BibitemOpen
  \bibfield  {author} {\bibinfo {author} {\bibfnamefont {P.}~\bibnamefont
  {Bruno}}, \bibinfo {author} {\bibfnamefont {V.~K.}\ \bibnamefont {Dugaev}},\
  and\ \bibinfo {author} {\bibfnamefont {M.}~\bibnamefont {Taillefumier}},\
  }\bibfield  {title} {\bibinfo {title} {Topological {Hall} effect and {Berry}
  phase in magnetic nanostructures.},\ }\href
  {https://doi.org/10.1103/PhysRevLett.93.096806} {\bibfield  {journal}
  {\bibinfo  {journal} {Phys. Rev. Lett.}\ }\textbf {\bibinfo {volume} {93}},\
  \bibinfo {pages} {096806} (\bibinfo {year} {2004})}\BibitemShut {NoStop}%
\bibitem [{\citenamefont {Bouaziz}\ \emph {et~al.}(2021)\citenamefont
  {Bouaziz}, \citenamefont {Ishida}, \citenamefont {Lounis},\ and\
  \citenamefont {Bl\"ugel}}]{Bouaziz:2021}%
  \BibitemOpen
  \bibfield  {author} {\bibinfo {author} {\bibfnamefont {J.}~\bibnamefont
  {Bouaziz}}, \bibinfo {author} {\bibfnamefont {H.}~\bibnamefont {Ishida}},
  \bibinfo {author} {\bibfnamefont {S.}~\bibnamefont {Lounis}},\ and\ \bibinfo
  {author} {\bibfnamefont {S.}~\bibnamefont {Bl\"ugel}},\ }\bibfield  {title}
  {\bibinfo {title} {Transverse transport in two-dimensional relativistic
  systems with nontrivial spin textures.},\ }\href
  {https://doi.org/10.1103/PhysRevLett.126.147203} {\bibfield  {journal}
  {\bibinfo  {journal} {Phys. Rev. Lett.}\ }\textbf {\bibinfo {volume} {126}},\
  \bibinfo {pages} {147203} (\bibinfo {year} {2021})}\BibitemShut {NoStop}%
\bibitem [{\citenamefont {Jan}(1957)}]{Jan:1957}%
  \BibitemOpen
  \bibfield  {author} {\bibinfo {author} {\bibfnamefont {J.-P.}\ \bibnamefont
  {Jan}},\ }in\ \href
  {https://doi.org/https://doi.org/10.1016/S0081-1947(08)60101-0} {\emph
  {\bibinfo {booktitle} {Galvamomagnetic and thermomagnetic effects in
  metals.}}},\ \bibinfo {series} {Solid State Physics}, Vol.~\bibinfo {volume}
  {5},\ \bibinfo {editor} {edited by\ \bibinfo {editor} {\bibfnamefont
  {F.}~\bibnamefont {Seitz}}\ and\ \bibinfo {editor} {\bibfnamefont
  {D.}~\bibnamefont {Turnbull}}}\ (\bibinfo  {publisher} {Academic Press},\
  \bibinfo {year} {1957})\ pp.\ \bibinfo {pages} {1--96}\BibitemShut {NoStop}%
\bibitem [{\citenamefont {Seemann}\ \emph {et~al.}(2011)\citenamefont
  {Seemann}, \citenamefont {Freimuth}, \citenamefont {Zhang}, \citenamefont
  {Bl\"ugel}, \citenamefont {Mokrousov}, \citenamefont {B\"urgler},\ and\
  \citenamefont {Schneider}}]{Seemann:2011}%
  \BibitemOpen
  \bibfield  {author} {\bibinfo {author} {\bibfnamefont {K.~M.}\ \bibnamefont
  {Seemann}}, \bibinfo {author} {\bibfnamefont {F.}~\bibnamefont {Freimuth}},
  \bibinfo {author} {\bibfnamefont {H.}~\bibnamefont {Zhang}}, \bibinfo
  {author} {\bibfnamefont {S.}~\bibnamefont {Bl\"ugel}}, \bibinfo {author}
  {\bibfnamefont {Y.}~\bibnamefont {Mokrousov}}, \bibinfo {author}
  {\bibfnamefont {D.~E.}\ \bibnamefont {B\"urgler}},\ and\ \bibinfo {author}
  {\bibfnamefont {C.~M.}\ \bibnamefont {Schneider}},\ }\bibfield  {title}
  {\bibinfo {title} {Origin of the planar {Hall} effect in nanocrystalline
  {${\mathrm{Co}}_{60}{\mathrm{Fe}}_{20}{\mathrm{B}}_{20}$}.},\ }\href
  {https://doi.org/10.1103/PhysRevLett.107.086603} {\bibfield  {journal}
  {\bibinfo  {journal} {Phys. Rev. Lett.}\ }\textbf {\bibinfo {volume} {107}},\
  \bibinfo {pages} {086603} (\bibinfo {year} {2011})}\BibitemShut {NoStop}%
\bibitem [{\citenamefont {Bouaziz}\ \emph {et~al.}(2016)\citenamefont
  {Bouaziz}, \citenamefont {Lounis}, \citenamefont {Bl\"ugel},\ and\
  \citenamefont {Ishida}}]{Juba:2016}%
  \BibitemOpen
  \bibfield  {author} {\bibinfo {author} {\bibfnamefont {J.}~\bibnamefont
  {Bouaziz}}, \bibinfo {author} {\bibfnamefont {S.}~\bibnamefont {Lounis}},
  \bibinfo {author} {\bibfnamefont {S.}~\bibnamefont {Bl\"ugel}},\ and\
  \bibinfo {author} {\bibfnamefont {H.}~\bibnamefont {Ishida}},\ }\bibfield
  {title} {\bibinfo {title} {Microscopic theory of the residual surface
  resistivity of {Rashba} electrons.},\ }\href
  {https://doi.org/10.1103/PhysRevB.94.045433} {\bibfield  {journal} {\bibinfo
  {journal} {Phys. Rev. B}\ }\textbf {\bibinfo {volume} {94}},\ \bibinfo
  {pages} {045433} (\bibinfo {year} {2016})}\BibitemShut {NoStop}%
\bibitem [{\citenamefont {Li}\ \emph {et~al.}(2020)\citenamefont {Li},
  \citenamefont {Xiao}, \citenamefont {Zou}, \citenamefont {Li}, \citenamefont
  {Zhang}, \citenamefont {Zeng}, \citenamefont {Zhou}, \citenamefont {Zhang},\
  and\ \citenamefont {Wu}}]{li:2020}%
  \BibitemOpen
  \bibfield  {author} {\bibinfo {author} {\bibfnamefont {Z.}~\bibnamefont
  {Li}}, \bibinfo {author} {\bibfnamefont {T.}~\bibnamefont {Xiao}}, \bibinfo
  {author} {\bibfnamefont {R.}~\bibnamefont {Zou}}, \bibinfo {author}
  {\bibfnamefont {J.}~\bibnamefont {Li}}, \bibinfo {author} {\bibfnamefont
  {Y.}~\bibnamefont {Zhang}}, \bibinfo {author} {\bibfnamefont
  {Y.}~\bibnamefont {Zeng}}, \bibinfo {author} {\bibfnamefont {M.}~\bibnamefont
  {Zhou}}, \bibinfo {author} {\bibfnamefont {J.}~\bibnamefont {Zhang}},\ and\
  \bibinfo {author} {\bibfnamefont {W.}~\bibnamefont {Wu}},\ }\bibfield
  {title} {\bibinfo {title} {Planar {Hall} effect in {PtSe$_{2}$}.},\ }\href
  {https://doi.org/10.1063/1.5133809} {\bibfield  {journal} {\bibinfo
  {journal} {J. Appl. Phys.}\ }\textbf {\bibinfo {volume} {127}},\ \bibinfo
  {pages} {054306} (\bibinfo {year} {2020})}\BibitemShut {NoStop}%
\bibitem [{\citenamefont {Togawa}\ \emph {et~al.}(2013)\citenamefont {Togawa},
  \citenamefont {Kousaka}, \citenamefont {Nishihara}, \citenamefont {Inoue},
  \citenamefont {akimitsu}, \citenamefont {Ovchinnikov},\ and\ \citenamefont
  {Kishine}}]{Togawa:2013}%
  \BibitemOpen
  \bibfield  {author} {\bibinfo {author} {\bibfnamefont {Y.}~\bibnamefont
  {Togawa}}, \bibinfo {author} {\bibfnamefont {Y.}~\bibnamefont {Kousaka}},
  \bibinfo {author} {\bibfnamefont {S.}~\bibnamefont {Nishihara}}, \bibinfo
  {author} {\bibfnamefont {K.}~\bibnamefont {Inoue}}, \bibinfo {author}
  {\bibfnamefont {J.}~\bibnamefont {akimitsu}}, \bibinfo {author}
  {\bibfnamefont {A.~S.}\ \bibnamefont {Ovchinnikov}},\ and\ \bibinfo {author}
  {\bibfnamefont {J.}~\bibnamefont {Kishine}},\ }\bibfield  {title} {\bibinfo
  {title} {Interlayer magnetoresistance due to chiral soliton lattice formation
  in hexagonal chiral magnet {CrNb$_3$S$_6$}.},\ }\href
  {https://link.aps.org/doi/10.1103/PhysRevLett.111.197204} {\bibfield
  {journal} {\bibinfo  {journal} {Phys. Rev. Lett.}\ }\textbf {\bibinfo
  {volume} {111}},\ \bibinfo {pages} {197204} (\bibinfo {year}
  {2013})}\BibitemShut {NoStop}%
\bibitem [{\citenamefont {Han}\ \emph {et~al.}(2017)\citenamefont {Han},
  \citenamefont {Zhang}, \citenamefont {Sapkota}, \citenamefont {Hao},
  \citenamefont {Ling}, \citenamefont {Du}, \citenamefont {Pi}, \citenamefont
  {Zhang}, \citenamefont {Mandrus},\ and\ \citenamefont {Zhang}}]{Hans:2017}%
  \BibitemOpen
  \bibfield  {author} {\bibinfo {author} {\bibfnamefont {H.}~\bibnamefont
  {Han}}, \bibinfo {author} {\bibfnamefont {L.}~\bibnamefont {Zhang}}, \bibinfo
  {author} {\bibfnamefont {D.}~\bibnamefont {Sapkota}}, \bibinfo {author}
  {\bibfnamefont {N.}~\bibnamefont {Hao}}, \bibinfo {author} {\bibfnamefont
  {L.}~\bibnamefont {Ling}}, \bibinfo {author} {\bibfnamefont {H.}~\bibnamefont
  {Du}}, \bibinfo {author} {\bibfnamefont {L.}~\bibnamefont {Pi}}, \bibinfo
  {author} {\bibfnamefont {C.}~\bibnamefont {Zhang}}, \bibinfo {author}
  {\bibfnamefont {D.~G.}\ \bibnamefont {Mandrus}},\ and\ \bibinfo {author}
  {\bibfnamefont {Y.}~\bibnamefont {Zhang}},\ }\bibfield  {title} {\bibinfo
  {title} {Tricritical point and phase diagram based on critical scaling in the
  monoaxial chiral helimagnet {Cr$_{1/3}$NbS$_2$}},\ }\href
  {https://doi.org/10.1103/PhysRevB.96.094439} {\bibfield  {journal} {\bibinfo
  {journal} {Phys. Rev. B}\ }\textbf {\bibinfo {volume} {96}},\ \bibinfo
  {pages} {094439} (\bibinfo {year} {2017})}\BibitemShut {NoStop}%
\bibitem [{\citenamefont {Bornstein}\ \emph {et~al.}(2015)\citenamefont
  {Bornstein}, \citenamefont {Chapman}, \citenamefont {Ghimire}, \citenamefont
  {Mandrus}, \citenamefont {Parker},\ and\ \citenamefont
  {Lee}}]{Bornstein:2015}%
  \BibitemOpen
  \bibfield  {author} {\bibinfo {author} {\bibfnamefont {A.~C.}\ \bibnamefont
  {Bornstein}}, \bibinfo {author} {\bibfnamefont {B.~J.}\ \bibnamefont
  {Chapman}}, \bibinfo {author} {\bibfnamefont {N.~J.}\ \bibnamefont
  {Ghimire}}, \bibinfo {author} {\bibfnamefont {D.~G.}\ \bibnamefont
  {Mandrus}}, \bibinfo {author} {\bibfnamefont {D.~S.}\ \bibnamefont
  {Parker}},\ and\ \bibinfo {author} {\bibfnamefont {M.}~\bibnamefont {Lee}},\
  }\bibfield  {title} {\bibinfo {title} {Out-of-plane spin-orientation
  dependent magnetotransport properties in the anisotropic helimagnet
  {Cr$_{1/3}$NbS$_{2}$}.},\ }\href {https://doi.org/10.1103/PhysRevB.91.184401}
  {\bibfield  {journal} {\bibinfo  {journal} {Phys. Rev. B}\ }\textbf {\bibinfo
  {volume} {91}},\ \bibinfo {pages} {184401} (\bibinfo {year}
  {2015})}\BibitemShut {NoStop}%
\bibitem [{\citenamefont {Maryenko}\ \emph {et~al.}(2017)\citenamefont
  {Maryenko}, \citenamefont {Mishchenko}, \citenamefont {Bahramy},
  \citenamefont {Ernst}, \citenamefont {Falson}, \citenamefont {Kozuka},
  \citenamefont {Tsukazaki}, \citenamefont {Nagaosa},\ and\ \citenamefont
  {Kawasaki}}]{Maryenko:2017}%
  \BibitemOpen
  \bibfield  {author} {\bibinfo {author} {\bibfnamefont {D.}~\bibnamefont
  {Maryenko}}, \bibinfo {author} {\bibfnamefont {A.~S.}\ \bibnamefont
  {Mishchenko}}, \bibinfo {author} {\bibfnamefont {M.~S.}\ \bibnamefont
  {Bahramy}}, \bibinfo {author} {\bibfnamefont {A.}~\bibnamefont {Ernst}},
  \bibinfo {author} {\bibfnamefont {J.}~\bibnamefont {Falson}}, \bibinfo
  {author} {\bibfnamefont {Y.}~\bibnamefont {Kozuka}}, \bibinfo {author}
  {\bibfnamefont {A.}~\bibnamefont {Tsukazaki}}, \bibinfo {author}
  {\bibfnamefont {N.}~\bibnamefont {Nagaosa}},\ and\ \bibinfo {author}
  {\bibfnamefont {M.}~\bibnamefont {Kawasaki}},\ }\bibfield  {title} {\bibinfo
  {title} {Observation of anomalous {Hall} effect in non-magnetic
  two-dimensional electron system.},\ }\href
  {https://doi.org/10.1038/ncomms14777} {\bibfield  {journal} {\bibinfo
  {journal} {Nat. Commun.}\ }\textbf {\bibinfo {volume} {8}},\ \bibinfo {pages}
  {14777} (\bibinfo {year} {2017})}\BibitemShut {NoStop}%
\bibitem [{\citenamefont {Onoda}\ \emph {et~al.}(2008)\citenamefont {Onoda},
  \citenamefont {Sugimoto},\ and\ \citenamefont {Nagaosa}}]{Onada:2008}%
  \BibitemOpen
  \bibfield  {author} {\bibinfo {author} {\bibfnamefont {S.}~\bibnamefont
  {Onoda}}, \bibinfo {author} {\bibfnamefont {N.}~\bibnamefont {Sugimoto}},\
  and\ \bibinfo {author} {\bibfnamefont {N.}~\bibnamefont {Nagaosa}},\
  }\bibfield  {title} {\bibinfo {title} {Quantum transport theory of anomalous
  electric, thermoelectric, and thermal {Hall} effects in ferromagnets.},\
  }\href {https://doi.org/10.1103/PhysRevB.77.165103} {\bibfield  {journal}
  {\bibinfo  {journal} {Phys. Rev. B}\ }\textbf {\bibinfo {volume} {77}},\
  \bibinfo {pages} {165103} (\bibinfo {year} {2008})}\BibitemShut {NoStop}%
\bibitem [{\citenamefont {Ishizuka}\ and\ \citenamefont
  {Nagaosa}(2018)}]{Ishizuka:2018}%
  \BibitemOpen
  \bibfield  {author} {\bibinfo {author} {\bibfnamefont {H.}~\bibnamefont
  {Ishizuka}}\ and\ \bibinfo {author} {\bibfnamefont {N.}~\bibnamefont
  {Nagaosa}},\ }\bibfield  {title} {\bibinfo {title} {Spin chirality induced
  skew scattering and anomalous {Hall} effect in chiral magnets},\ }\href
  {https://doi.org/10.1126/sciadv.aap9962} {\bibfield  {journal} {\bibinfo
  {journal} {Science Advances}\ }\textbf {\bibinfo {volume} {4}},\ \bibinfo
  {pages} {eaap9962} (\bibinfo {year} {2018})}\BibitemShut {NoStop}%
\bibitem [{\citenamefont {Li}\ \emph {et~al.}(2013)\citenamefont {Li},
  \citenamefont {Kanazawa}, \citenamefont {Yu}, \citenamefont {Tsukazaki},
  \citenamefont {Kawasaki}, \citenamefont {Ichikawa}, \citenamefont {Jin},
  \citenamefont {Kagawa},\ and\ \citenamefont {Tokura}}]{Li:2013}%
  \BibitemOpen
  \bibfield  {author} {\bibinfo {author} {\bibfnamefont {Y.}~\bibnamefont
  {Li}}, \bibinfo {author} {\bibfnamefont {N.}~\bibnamefont {Kanazawa}},
  \bibinfo {author} {\bibfnamefont {X.~Z.}\ \bibnamefont {Yu}}, \bibinfo
  {author} {\bibfnamefont {A.}~\bibnamefont {Tsukazaki}}, \bibinfo {author}
  {\bibfnamefont {M.}~\bibnamefont {Kawasaki}}, \bibinfo {author}
  {\bibfnamefont {M.}~\bibnamefont {Ichikawa}}, \bibinfo {author}
  {\bibfnamefont {X.~F.}\ \bibnamefont {Jin}}, \bibinfo {author} {\bibfnamefont
  {F.}~\bibnamefont {Kagawa}},\ and\ \bibinfo {author} {\bibfnamefont
  {Y.}~\bibnamefont {Tokura}},\ }\bibfield  {title} {\bibinfo {title} {Robust
  formation of skyrmions and topological {Hall} effect anomaly in epitaxial
  thin films of {MnSi}.},\ }\href
  {https://doi.org/10.1103/PhysRevLett.110.117202} {\bibfield  {journal}
  {\bibinfo  {journal} {Phys. Rev. Lett.}\ }\textbf {\bibinfo {volume} {110}},\
  \bibinfo {pages} {117202} (\bibinfo {year} {2013})}\BibitemShut {NoStop}%
\bibitem [{\citenamefont {Paterson}\ \emph {et~al.}(2020)\citenamefont
  {Paterson}, \citenamefont {Tereshchenko}, \citenamefont {Nakayama},
  \citenamefont {Kousaka}, \citenamefont {Kishine}, \citenamefont {McVitie},
  \citenamefont {Ovchinnikov}, \citenamefont {Proskurin},\ and\ \citenamefont
  {Togawa}}]{Paterson:2020}%
  \BibitemOpen
  \bibfield  {author} {\bibinfo {author} {\bibfnamefont {G.~W.}\ \bibnamefont
  {Paterson}}, \bibinfo {author} {\bibfnamefont {A.~A.}\ \bibnamefont
  {Tereshchenko}}, \bibinfo {author} {\bibfnamefont {S.}~\bibnamefont
  {Nakayama}}, \bibinfo {author} {\bibfnamefont {Y.}~\bibnamefont {Kousaka}},
  \bibinfo {author} {\bibfnamefont {J.}~\bibnamefont {Kishine}}, \bibinfo
  {author} {\bibfnamefont {S.}~\bibnamefont {McVitie}}, \bibinfo {author}
  {\bibfnamefont {A.~S.}\ \bibnamefont {Ovchinnikov}}, \bibinfo {author}
  {\bibfnamefont {I.}~\bibnamefont {Proskurin}},\ and\ \bibinfo {author}
  {\bibfnamefont {Y.}~\bibnamefont {Togawa}},\ }\bibfield  {title} {\bibinfo
  {title} {Tensile deformations of the magnetic chiral soliton lattice probed
  by {Lorentz} transmission electron microscopy.},\ }\href
  {https://doi.org/10.1103/PhysRevB.101.184424} {\bibfield  {journal} {\bibinfo
   {journal} {Phys. Rev. B}\ }\textbf {\bibinfo {volume} {101}},\ \bibinfo
  {pages} {184424} (\bibinfo {year} {2020})}\BibitemShut {NoStop}%
\bibitem [{\citenamefont {Tsuruta}\ \emph {et~al.}(2016)\citenamefont
  {Tsuruta}, \citenamefont {Mito}, \citenamefont {Deguchi}, \citenamefont
  {Kishine}, \citenamefont {Kousaka}, \citenamefont {Akimitsu},\ and\
  \citenamefont {Inoue}}]{Tsuruta:2016}%
  \BibitemOpen
  \bibfield  {author} {\bibinfo {author} {\bibfnamefont {K.}~\bibnamefont
  {Tsuruta}}, \bibinfo {author} {\bibfnamefont {M.}~\bibnamefont {Mito}},
  \bibinfo {author} {\bibfnamefont {H.}~\bibnamefont {Deguchi}}, \bibinfo
  {author} {\bibfnamefont {J.}~\bibnamefont {Kishine}}, \bibinfo {author}
  {\bibfnamefont {Y.}~\bibnamefont {Kousaka}}, \bibinfo {author} {\bibfnamefont
  {J.}~\bibnamefont {Akimitsu}},\ and\ \bibinfo {author} {\bibfnamefont
  {K.}~\bibnamefont {Inoue}},\ }\bibfield  {title} {\bibinfo {title} {Phase
  diagram of the chiral magnet {Cr$_{1/3}$NbS$_{2}$} in a magnetic field.},\
  }\href {https://doi.org/10.1103/PhysRevB.93.104402} {\bibfield  {journal}
  {\bibinfo  {journal} {Phys. Rev. B}\ }\textbf {\bibinfo {volume} {93}},\
  \bibinfo {pages} {104402} (\bibinfo {year} {2016})}\BibitemShut {NoStop}%
\bibitem [{dat()}]{data_portal}%
  \BibitemOpen
  \href@noop {} {}\bibinfo {note} {See
  http://wrap.warwick.ac.uk/162181/}\BibitemShut {NoStop}%
\bibitem [{\citenamefont {M\"uller}\ \emph {et~al.}(2019)\citenamefont
  {M\"uller}, \citenamefont {Hoffmann}, \citenamefont {Di\ss{}elkamp},
  \citenamefont {Sch\"urhoff}, \citenamefont {Mavros}, \citenamefont
  {Sallermann}, \citenamefont {Kiselev}, \citenamefont {J\'onsson},\ and\
  \citenamefont {Bl\"ugel}}]{Gideon:2019}%
  \BibitemOpen
  \bibfield  {author} {\bibinfo {author} {\bibfnamefont {G.~P.}\ \bibnamefont
  {M\"uller}}, \bibinfo {author} {\bibfnamefont {M.}~\bibnamefont {Hoffmann}},
  \bibinfo {author} {\bibfnamefont {C.}~\bibnamefont {Di\ss{}elkamp}}, \bibinfo
  {author} {\bibfnamefont {D.}~\bibnamefont {Sch\"urhoff}}, \bibinfo {author}
  {\bibfnamefont {S.}~\bibnamefont {Mavros}}, \bibinfo {author} {\bibfnamefont
  {M.}~\bibnamefont {Sallermann}}, \bibinfo {author} {\bibfnamefont {N.~S.}\
  \bibnamefont {Kiselev}}, \bibinfo {author} {\bibfnamefont {H.}~\bibnamefont
  {J\'onsson}},\ and\ \bibinfo {author} {\bibfnamefont {S.}~\bibnamefont
  {Bl\"ugel}},\ }\bibfield  {title} {\bibinfo {title} {Spirit: Multifunctional
  framework for atomistic spin simulations.},\ }\href
  {https://doi.org/10.1103/PhysRevB.99.224414} {\bibfield  {journal} {\bibinfo
  {journal} {Phys. Rev. B}\ }\textbf {\bibinfo {volume} {99}},\ \bibinfo
  {pages} {224414} (\bibinfo {year} {2019})}\BibitemShut {NoStop}%
\end{thebibliography}%


\begin{figure*}[tb!]
\centering
\includegraphics[width=0.95\textwidth]{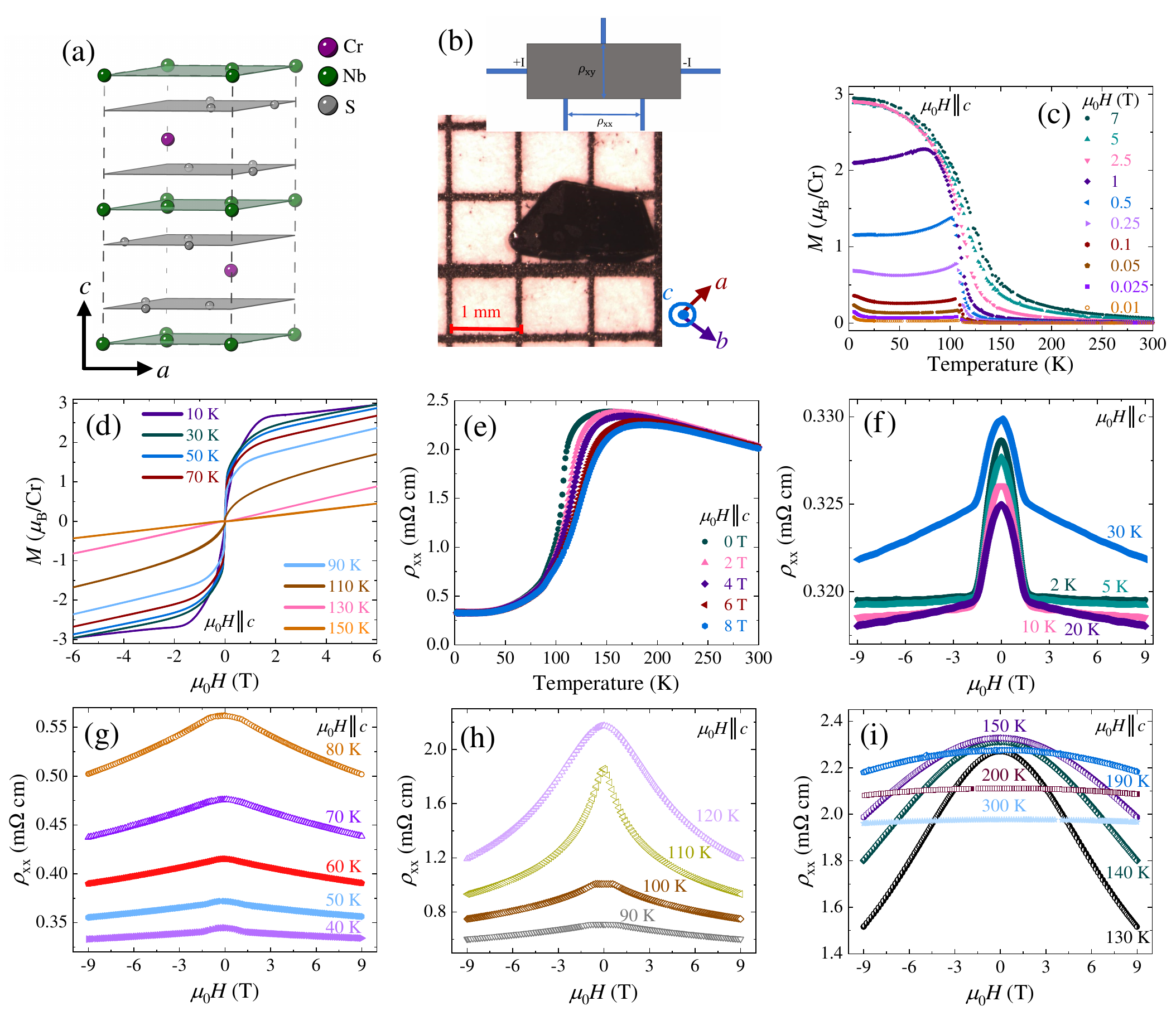}
\caption{(Color online) (a) Crystal structure of Cr$_{1/3}$NbS$_{2}$. The purple, green and gray spheres denote the chromium, niobium and sulfur atoms, respectively. The dashed line indicates the unit cell. (b) A single crystal of Cr$_{1/3}$NbS$_{2}$ grown by chemical vapor transport that was used for all the measurements presented in this work, along with a schematic of the 5-point wire configuration used for the longitudinal and Hall resistivity measurements. (c) Magnetization versus temperature curves for Cr$_{1/3}$NbS$_{2}$ for fields of up to 7~T applied perpendicular to the platelet. (d) Magnetization as a function of field applied parallel to the $c$-axis at several temperatures. (e) Temperature dependence of the longitudinal resistivity $\rho_{\mathrm{xx}}$ in several magnetic fields applied parallel to the $c$-axis. Magnetoresistances of Cr$_{1/3}$NbS$_{2}$ between -9 and 9~T at (f) 2 to 30~K, (g) 40 to 80~K, (h) 90 to 120~K and (i) 130 to 300~K.}
\label{FIG: Mag and Res}
\label{FIG: Crystal structure}
\end{figure*}

\begin{figure*}[tb]
\centering
\includegraphics[width=0.9\textwidth]{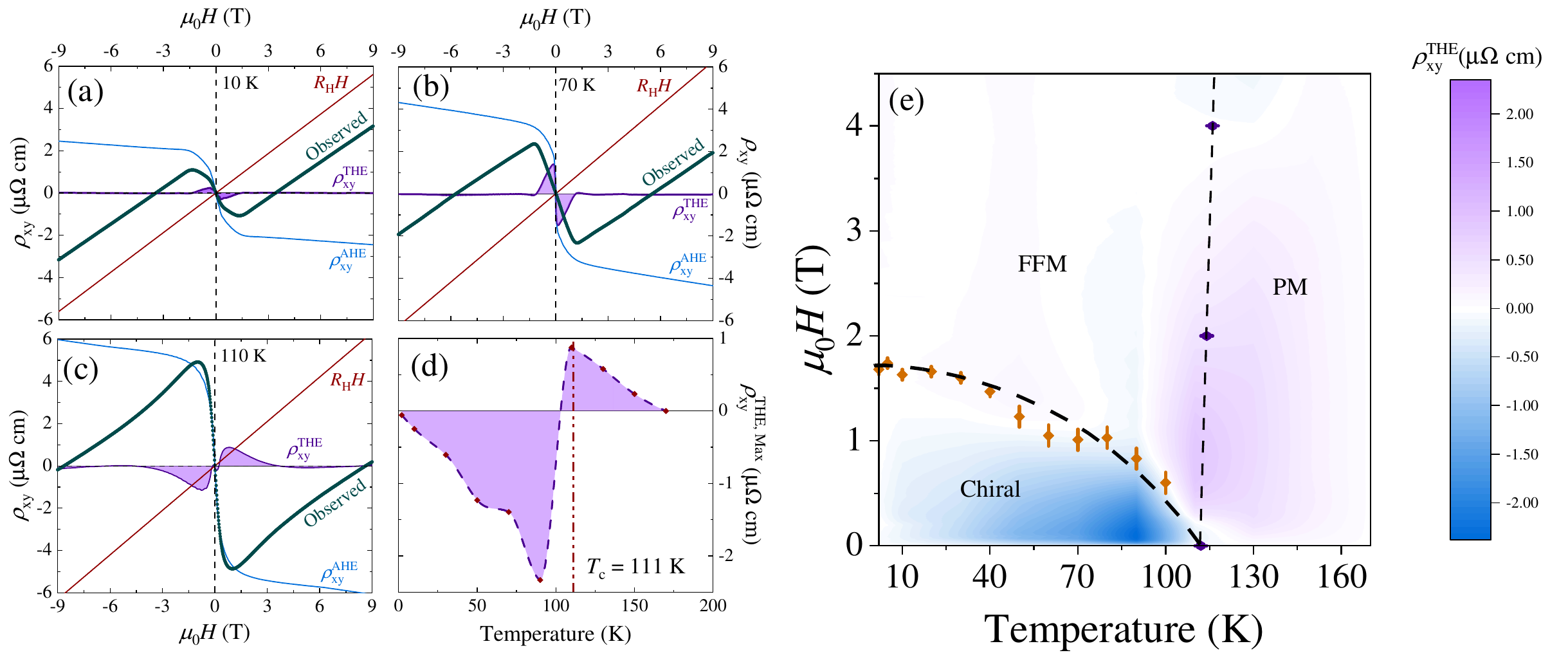}
\caption{(Color online) Observed $\rho_{\mathrm{xy}}$ (green) as a function of applied magnetic field at (a) 10, (b) 70 and (c) 110~K with the calculated contributions of $\rho_{\mathrm{xy}}^{\mathrm{OHE}}$ (red), $\rho_{\mathrm{xy}}^{\mathrm{AHE}}$ (blue) and $\rho_{\mathrm{xy}}^{\mathrm{THE}}$ (purple). (d) The maximum $\rho_{\mathrm{xy}}^{\mathrm{THE}}$ as a function of temperature. (e) Contour map of $\rho_{\mathrm{xy}}^{\mathrm{THE}}$ as function of field and temperature. The purple points indicate the position of the transition from resistivity as seen in Fig.~\ref{FIG: Mag and Res}(b). The phase boundaries are approximated from the $\rho_{\mathrm{xx}}$ vs temperature (purple) and field (orange) data shown in Fig.~\ref{FIG: Mag and Res}. The dashed lines are estimations of phase boundaries in Cr$_{1/3}$NbS$_{2}$ marking out the chiral, FFM and paramagnetic (PM) phases.}
\label{FIG: Hall analysis}
\end{figure*}

\end{document}


\title{Supplemental Material: Giant topological and planar Hall effect in Cr$_{1/3}$NbS$_{2}$}

\author{D. A. Mayoh}
\email[]{d.mayoh.1@warwick.ac.uk}
\affiliation{Physics Department, University of Warwick, Coventry, CV4 7AL, United Kingdom}

\author{J. Bouaziz}
\affiliation{Physics Department, University of Warwick, Coventry, CV4 7AL, United Kingdom}

\author{A. E. Hall}
\affiliation{Physics Department, University of Warwick, Coventry, CV4 7AL, United Kingdom}

\author{J. B. Staunton}
\affiliation{Physics Department, University of Warwick, Coventry, CV4 7AL, United Kingdom}

\author{M. R. Lees}
\affiliation{Physics Department, University of Warwick, Coventry, CV4 7AL, United Kingdom}

\author{G. Balakrishnan}
\email[]{g.balakrishnan@warwick.ac.uk}
\affiliation{Physics Department, University of Warwick, Coventry, CV4 7AL, United Kingdom}

\maketitle

\section{Laue diffraction}

\begin{figure}[h]
\centering
\includegraphics[width=0.5\textwidth]{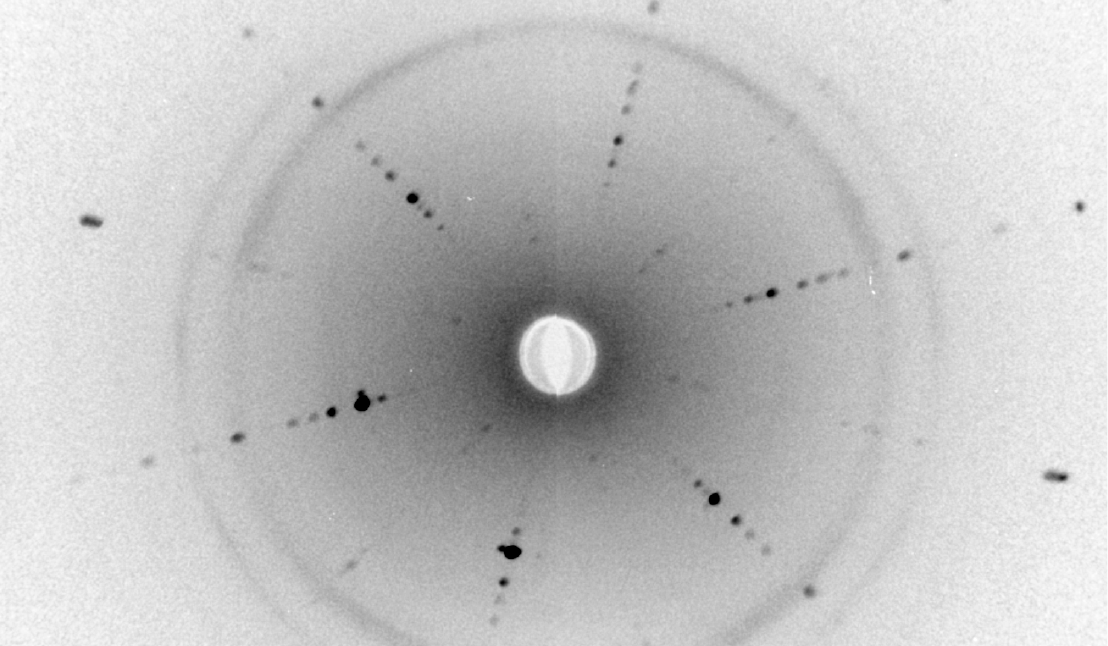}
\caption{X-ray Laue diffraction pattern of a platelet of Cr$_{1/3}$NbS$_{2}$ orientated along the [001] direction. The sample was orientated so that the x-rays where incident perpendicular to the plane of platelet.}
\label{FIG: Laue supplement}
\end{figure}

A Laue diffraction pattern of a Cr$_{1/3}$NbS$_{2}$ crystal with the x-ray beam perpendicular to the plane of the platelet is shown in Fig.~\ref{FIG: Laue supplement}. The observed pattern shows the crystal is orientated such that the $c$ axis is perpendicular to the plane of the platelet implying that the $ab$ plane lies within the plane of the platelet. This sample was used for all the measurements presented in this work. 

\section{Further Conventional and Planar Hall effect Data}

\begin{figure}[tb!]
\centering
\includegraphics[width=\textwidth]{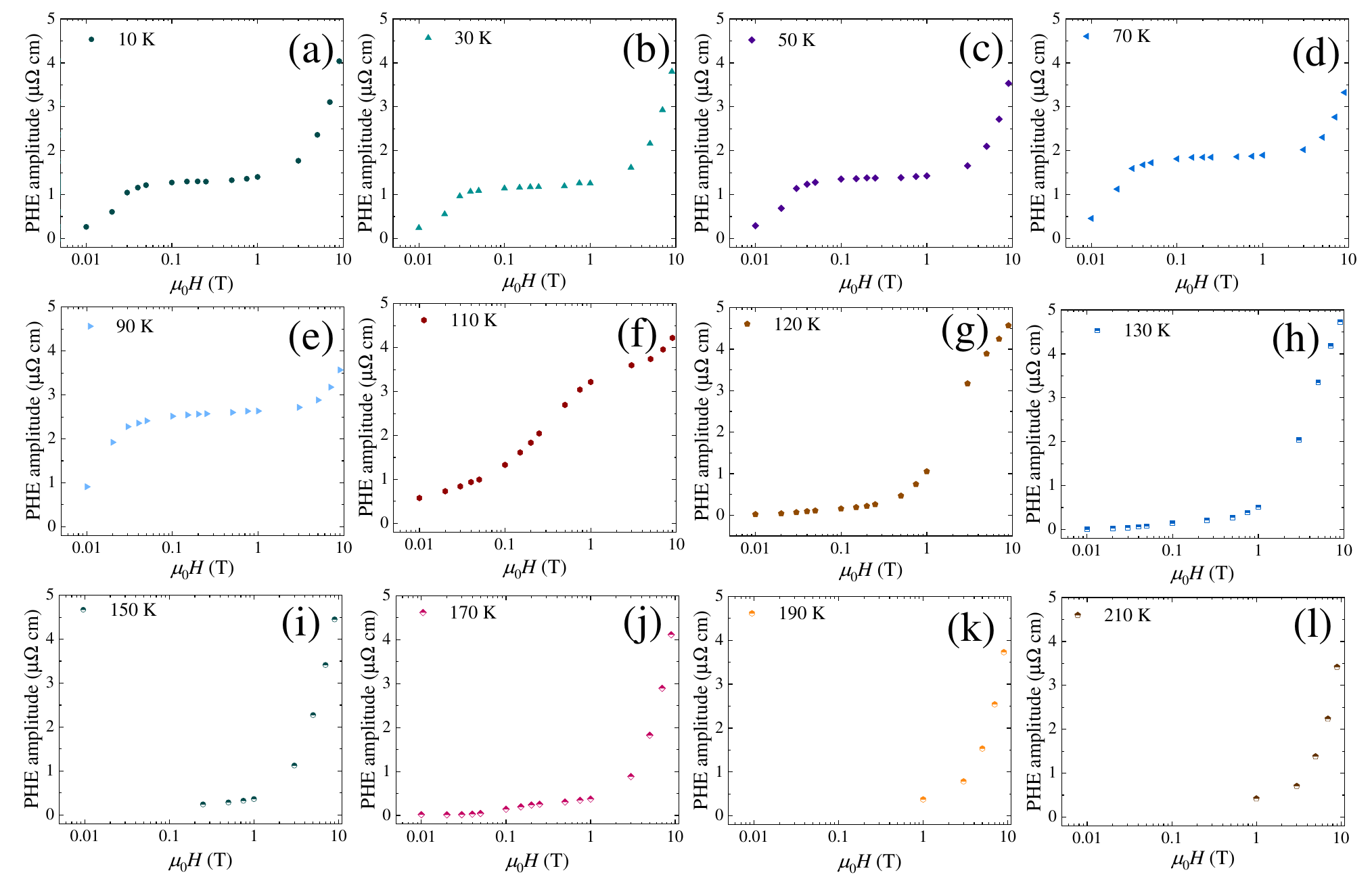}
\caption{Field dependence of the PHE amplitude in Cr$_{1/3}$NbS$_{2}$ at (a) 10, (b) 30, (c) 50, (d) 70, (e) 90, (f) 110, (g) 120, (h) 130, (i) 150, (j) 170, (k) 190, and (l) 210~K.}
\label{FIG: PHE supplement}
\end{figure}

The planar Hall effect (PHE) amplitude as a function of field for Cr$_{1/3}$NbS$_{2}$ at several temperatures is shown in Fig.~\ref{FIG: PHE supplement}. The PHE increases rapidly in low-fields at temperatures below the magnetic transition due to the tilted chiral soliton lattice (CSL). Above the magnetic transition in high fields, a finite planar Hall amplitude persists due to a ferromagnetic contribution [see Fig.~1(c)] to the magnetisation that is tilted out of plane.

\begin{figure}[tb!]
\centering
\includegraphics[width=\textwidth]{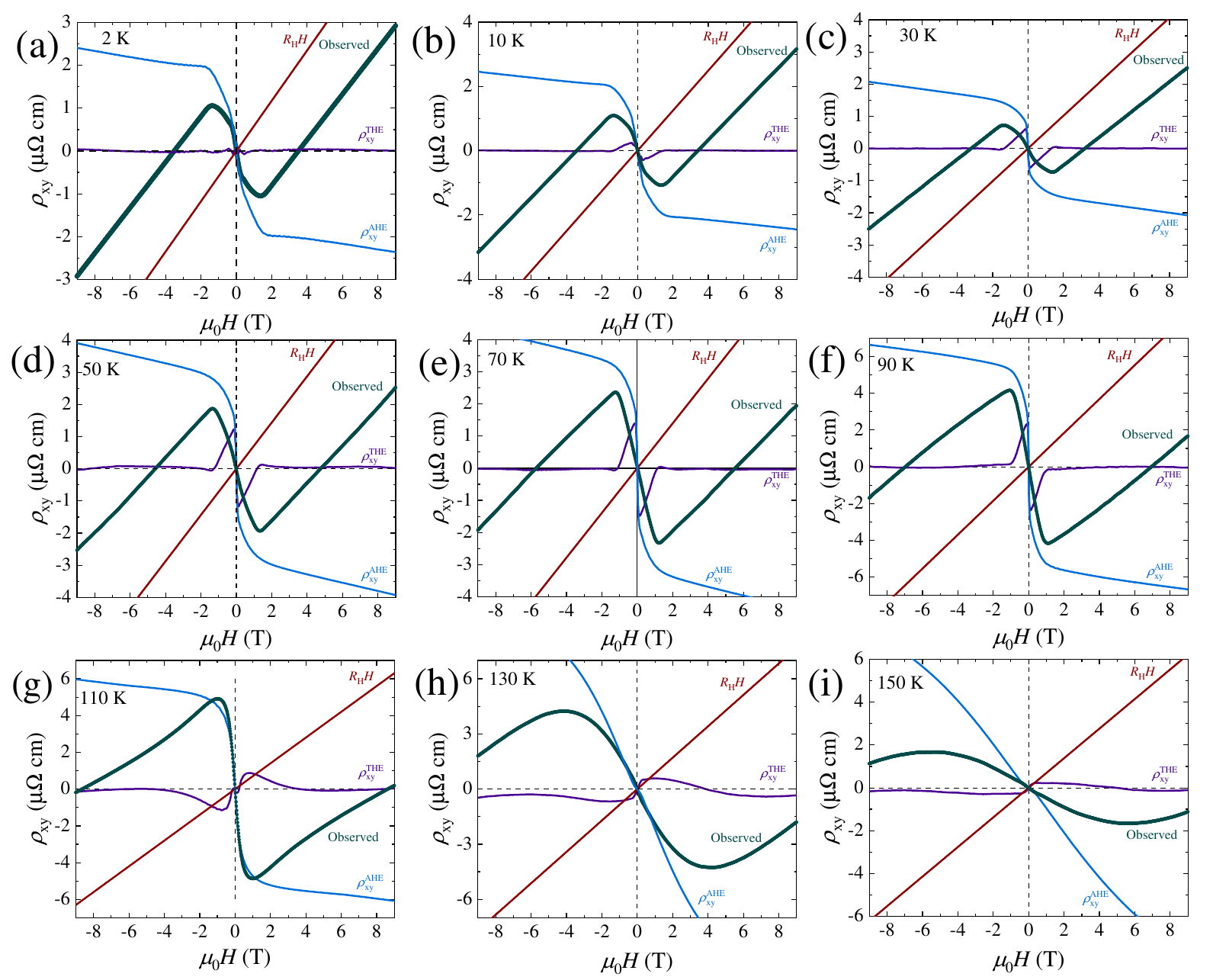}
\caption{Field dependence of the observed Hall effect signal (green), OHE (red), AHE (blue) and THE (purple) at (a) 2 (b) 10, (c) 30, (d) 50, (e) 70, (f) 90, (g) 110, (h) 130, and (i) 150~K.}
\label{FIG: Hall effect supplement}
\end{figure}

The conventional Hall effect as a function of field at several temperatures is shown in Fig.~\ref{FIG: Hall effect supplement}.

\bibliography{CrNb3S6_References.bib}